\documentclass[12pt]{article}
\pdfoutput=1

\usepackage{cite}

\usepackage[nointlimits,reqno]{amsmath}
\usepackage{amssymb}
\numberwithin{equation}{section} 

\usepackage{graphicx}
\usepackage{adjustbox}
\usepackage{paralist, tabularx}
\usepackage[toc,page]{appendix}

\usepackage{subcaption}
\usepackage{float}
\usepackage{caption}
\usepackage{luty}

\newcommand{\met}{\Sla{E}_{\rm T}}
\newcommand{\pt}{\ensuremath{p_{\rm T}}}

\begin{document}

\newcommand{\norm}[1]{|\!\gap|#1|\!\gap|}
\newcommand{\aavg}[1]{\avg{\!\avg{#1}\!}}

\begin{titlepage}

\title{Searching for Additional Higgs Bosons\\
via Higgs Cascades}

\author{Christina Gao,\ \ 
Markus A. Luty,\ \ 
Michael Mulhearn,\ \ 
\\
Nicol\'as A. Neill,\ \ 
Zhangqier Wang}

\address{Physics Department, University of California, Davis\\
Davis, California 95616}

\begin{abstract}
The discovery of a 125~GeV Higgs boson at the Large Hadron Collider strongly motivates direct searches for additional Higgs bosons.
In a type I two Higgs doublet model there is a large region of parameter space at $\tan\be \gsim 5$ that is currently unconstrained experimentally.
We show that the process $gg \to H \to A Z \to ZZh$ can probe this region,
and can be the discovery mode for an extended Higgs sector at the LHC.
We analyze 9 promising decay modes for the $ZZh$ state, and we find that
the most sensitive final states are $\ell\ell\ell\ell bb$,
$\ell\ell jjbb$, $\ell\ell\nu\nu\ga\ga$ and $\ell\ell\ell\ell +{}$missing energy.
\end{abstract}
\end{titlepage}

\section{Introduction}
The discovery of a Higgs-like 125 GeV particle at the Large Hadron
Collider (LHC) began a new era in elementary particle
physics \cite{Aad:2012tfa,Chatrchyan:2012xdj}.
The experiments have moved rapidly from discovery to 
precision measurements of the properties of the new particle.  
We now have strong
evidence that the 125 GeV particle is a scalar, and we know that its
couplings agree with the couplings of the Standard Model (SM) Higgs at
the 15\% level \cite{atlas3,cms2}.  
This implies that electroweak symmetry breaking is
(at least mostly) a weakly-coupled phenomenon.

As a direct result of this,
for the first time in the history of particle physics we have an
experimentally complete theory that can be consistently extrapolated
to energies far above the TeV scale.  This is a tremendous achievement,
but it also raises the question of whether there is any additional
new physics to be found at the LHC.  
The existence of an elementary scalar particle
brings to the fore the problem of naturalness of the standard model,
and strongly motivates new physics at the weak scale accessible to the
LHC.  In particular, extensions of the standard model that address the
naturalness problem generally require an extension of the Higgs
sector (for example SUSY or PNGB Higgs models).
Conversely, if the weak scale is unnatural and instead has an
anthropic explanation, we do not expect any additional light scalars.
Searching for additional Higgs bosons is therefore an
important part of the program of probing naturalness 
at the LHC.

There are many candidate models for physics beyond the standard model,
but the absence of any signal has ruled out the simplest and most natural models.
This motivates searching as widely as possible for new physics,
and suggests a phenomenological approach of looking
for signals based on plausible event topologies.  It is important not to
miss any possible signal, and at a hadron collider like the LHC one
has to know exactly what to look for to find new physics.  Simplified
models that contain the minimum degrees of freedom relevant for a
particular type of signal are useful tools for generating signal
events and interpreting the results of searches
\cite{Alves:2011wf}.  This is the philosophy adopted in the present
paper.

Our simplified model for additional Higgs bosons is the two Higgs
doublet model (2HDM).  
This is the simplest model that has an additional source of electroweak symmetry breaking 
beyond the Standard Model.  
In this paper, we study the Higgs cascade process $gg \to H \to ZA \to ZZh$.
We demonstrate that this can be important in a type I 2HDM for $\tan\be \gsim 5$.
Of course, the masses must be such that the decay is allowed, and we must have
$m_A < 2 m_t$, otherwise the decay $A \to t\bar{t}$ dominates and reduces our
signal.
Our search is therefore sensitive in the mass region
\[
215~\GeV < m_A < 355~\GeV,
\qquad
m_H > m_A + 90~\GeV.
\]
This is a region of parameter space where other direct
searches are insensitive.
We demonstrate that the searches we
consider can be more sensitive than indirect constraints from Higgs
coupling measurements, and in fact can be the discovery mode for
new Higgs bosons.
We analyzed 9 different final states for the
$ZZh$ decays that we believe are the most promising.  
These are listed in Table 1.
We use the notation $ZZh \to (\ell^+ \ell^- )(\nu\nu)(b\bar{b})$
to indicate $Z \to \ell^+ \ell^- $, $Z \to \nu\nu$, $h \to b\bar{b}$, {\it etc.},
where $\ell = e, \mu$.  
We find that the most sensitive final state
is the ``golden mode'' $\ell^+ \ell^- \ell^+ \ell^-  b \bar{b}$, 
which is essentially background-free, and can lead to a $5\si$
discovery of our our benchmark model at $66~$fb$^{-1}$ at the
14~TeV LHC.
Perhaps more surprisingly, we also find that the final state
$\ell^+ \ell^-  jjb\bar{b}$ is also sensitive.
This mode has large background, and a careful treatment of background
uncertainty is needed to draw any conclusion.
We obtain a $3\si$ sensitivity with $300$~fb$^{-1}$ with a cut-and-count
analysis, and find that a boosted decision tree technique may be able
to boost the significance above $5\si$.
We also find that the final states $(\ell^+ \ell^- )(\nu\nu)(\ga\ga)$
and $\ell^+ \ell^- \ell^+ \ell^- + \met$ are sensitive at the
$3\si$ and $2\si$ level, respectively.
Searches for $ZZh$ are therefore very
strongly motivated in the current run of the 14~TeV~LHC.

Because several of the searches are rate-limited, the 
high-luminosity LHC (3000~fb$^{-1}$ at 14~TeV) is very effective
in exploring the $ZZh$ signal.
We find that all five of the modes listed above
can give a $5\si$ signal for our benchmark model,
illustrating the possibility of studying the signal in multiple channels.

This paper is organized as follows.  In Section 2, we briefly review
the 2HDM and explain why the $ZZh$ final state is a sensitive probe
when in a type I 2HDM at large $\tan\be$.  In
Section 3, we report our analysis of experimental searches in a number
of final states and compute the significance for the 14~\TeV~LHC.
Section 4 contains our outlook and conclusions.

\begin{table}[H]
\centering
\begin{adjustbox}{max width=\textwidth}
\begin{tabular}{|l|l|l|}
\hline\hline
$ZZh$ decay modes & Comments & Significance\\ \hline
$(\ell^+ \ell^- )(\ell^+ \ell^- )(b\bar{b})$
& clean, ideal for reconstruction& $11\si$\\ \hline
$(\ell^+ \ell^- )(jj)(b\bar{b})$
& large signal and background& $3.6\si$\\\hline
$(\ell^+ \ell^- )(\nu\nu)(b\bar{b})$ 
& overwhelmed by $t\bar{t}$ background& small\\\hline
$(\ell^+ \ell^- )(jj)(\tau_h\tau_h)$ 
&  overwhelmed by jet-faked $\tau$ background& $0.7\si$\\ \hline
$(\ell^+ \ell^- )(\nu\nu)(\tau_h\tau_h)$ 
& not enough signal yield& small\\ \hline
$(\ell^+ \ell^- )(jj)(\ga \ga)$ 
& relatively clean, small signal cross section& $1.7\si$\\\hline
$(\nu\nu)(jj)(\ga \ga)$ 
& hard to reconstruct& $0.5\si$\\\hline
$(\ell^+ \ell^- )(\nu\nu)(\ga \ga)$ 
& clean but very small cross section& $2.8\si$\\ \hline
$\ell^+ \ell^- \ell^+ \ell^- +\met$
& relatively clean after hard $\met$ cut& $2.1\si$\\
\hline\hline
\end{tabular}
\end{adjustbox}
\caption{Final states of $ZZh$ considered in this paper, with comments
and signal significance for 14~\TeV\ LHC with integrated luminosity of 300~fb$^{-1}$.}
\label{zzh_finalstates}
\end{table}

\section{Additional Higgs Bosons in the Two Higgs Doublet Model}
In this section we briefly review the experimental constraints on the two
Higgs doublet model (2HDM) and motivate our search.  The 2HDM extends
the SM by adding an additional Higgs doublet.
It is by now a textbook topic \cite{Gunion:1989we}.

There are many parameters in a general 2HDM, but only a small number
of them are the most important for phenomenology.  One important
parameter of the model is the ratio of the VEVs of the two Higgs
doublets $H_{1,2}$, conventionally parametrized by $\tan\be = v_2/
v_1$, with $v = \sqrt{v_1^2 + v_2^2} = 246~\GeV$.  Another important
parameter is the mixing angle $\al$ that defines the mass eigenstates
of the neutral CP even scalar states:
\[
\begin{pmatrix} H^0 \\ h^0 \end{pmatrix}
= \begin{pmatrix} \cos\al & \sin\al \\ -\sin\al & \cos\al \end{pmatrix}
\begin{pmatrix} H_1^0 \\ H_2^0 \end{pmatrix}.
\]
For the interactions of the Higgs bosons with gauge bosons and other
Higgs bosons, there is no natural choice for $H_{1,2}$, so $\al$ and
$\be$ are not separately meaningful.  What is meaningful is the mixing
angle between the mass basis $(h^0, H^0)$ and the basis where one of
the fields has vanishing VEV.  This ``Higgs angle'' is given by $\theta_H
= \be - \al + \pi/2$.  (The $\pi/2$ is due to unfortunate standard
conventions.)  The coupling of all of the Higgs bosons to vector
bosons is therefore determined by $\be - \al$.  Some examples relevant
for our work are
\[
\begin{split}
AhZ & \propto \cos(\be - \al),
\qquad
AHZ  \propto \sin(\be - \al),
\\
hZZ & \propto \sin(\be - \al),
\qquad
HZZ  \propto \cos(\be - \al).
\end{split}
\]
The fact that the $hZZ$ coupling is modified for $\cos(\be - \al) \ne
0$ is an important constraint on this model.  In the ``alignment
limit'' $\cos(\be - \al) \to 0$ the light mass eigenstate $h^0$ is
solely responsible for electroweak symmetry breaking, and so its
couplings to all states are standard-model like.

The parameters $\al$ and $\be$ become separately meaningful when we
consider Yukawa couplings between the Higgs fields and fermions, which
pick out particular linear combinations of $H_{1,2}$.  The most
well-studied possibilities are so-called type I models where $H_2$
couples to all fermions, and type II models where $H_1$ couples to
leptons and down-type quarks, and $H_2$ couples to up-type quarks.
Type II models have received the most attention because the minimal
supersymmetric standard model (MSSM) Higgs sector is type II, but
extensions of the MSSM can have have an effective type I Higgs sector
at low energies \cite{Galloway:2013dma,Chang:2014ida,Alves:2012fx}.
There are other possibilities
beyond type I and II, but these are sufficient to illustrate the
physics being studied here.

\begin{figure}[h]
\centering
\begin{subfigure}[b]{0.45\textwidth}
\includegraphics[width=\textwidth]{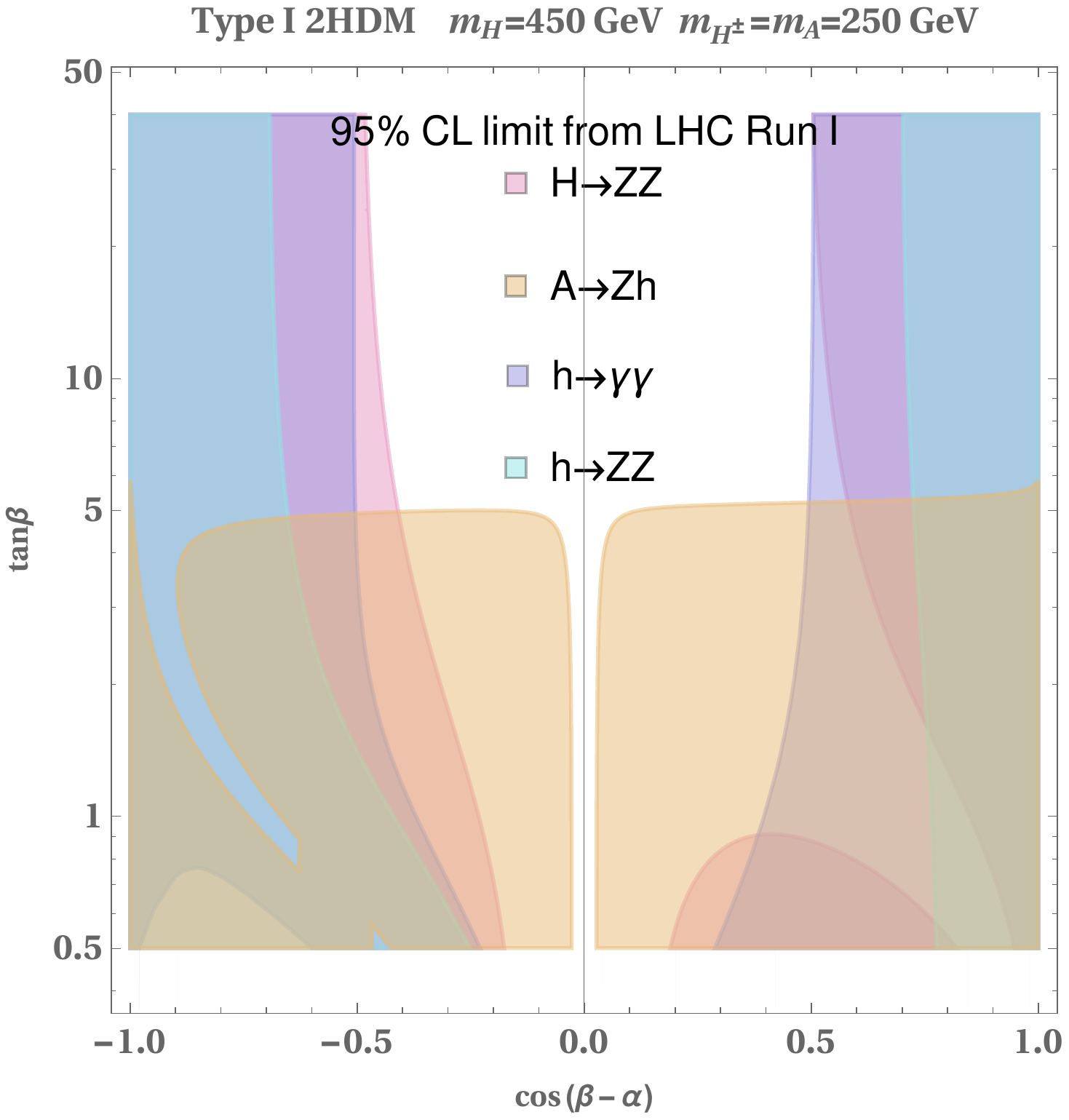}
\caption{}
\label{expbounds}
\end{subfigure}
\begin{subfigure}[b]{0.45\textwidth}
\includegraphics[width=\textwidth]{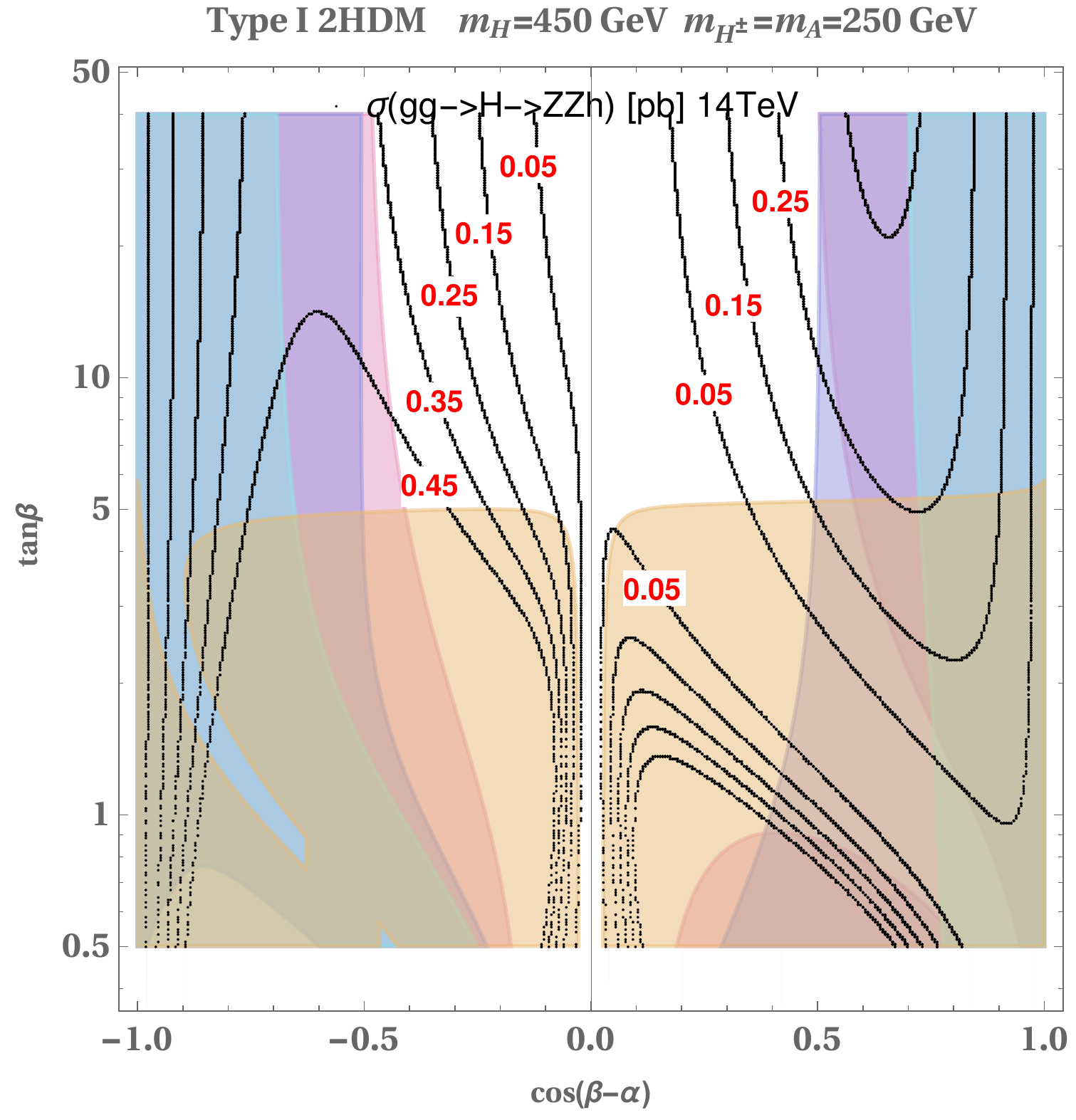}
\caption{}\label{zzh_b}
\end{subfigure}
\caption{Status of type I 2HDM after the 8~\TeV  run of the LHC
for $m_A = m_{H^\pm} = 250~\GeV$, $m_H = 450~\GeV$.
Plotted on the right are the values of the $\si \times \text{BR}$ for 
$gg \to H \to AZ \to ZZh$.}\label{zzh}
\end{figure}

\begin{figure}
\centering
\includegraphics[width=0.45\textwidth]{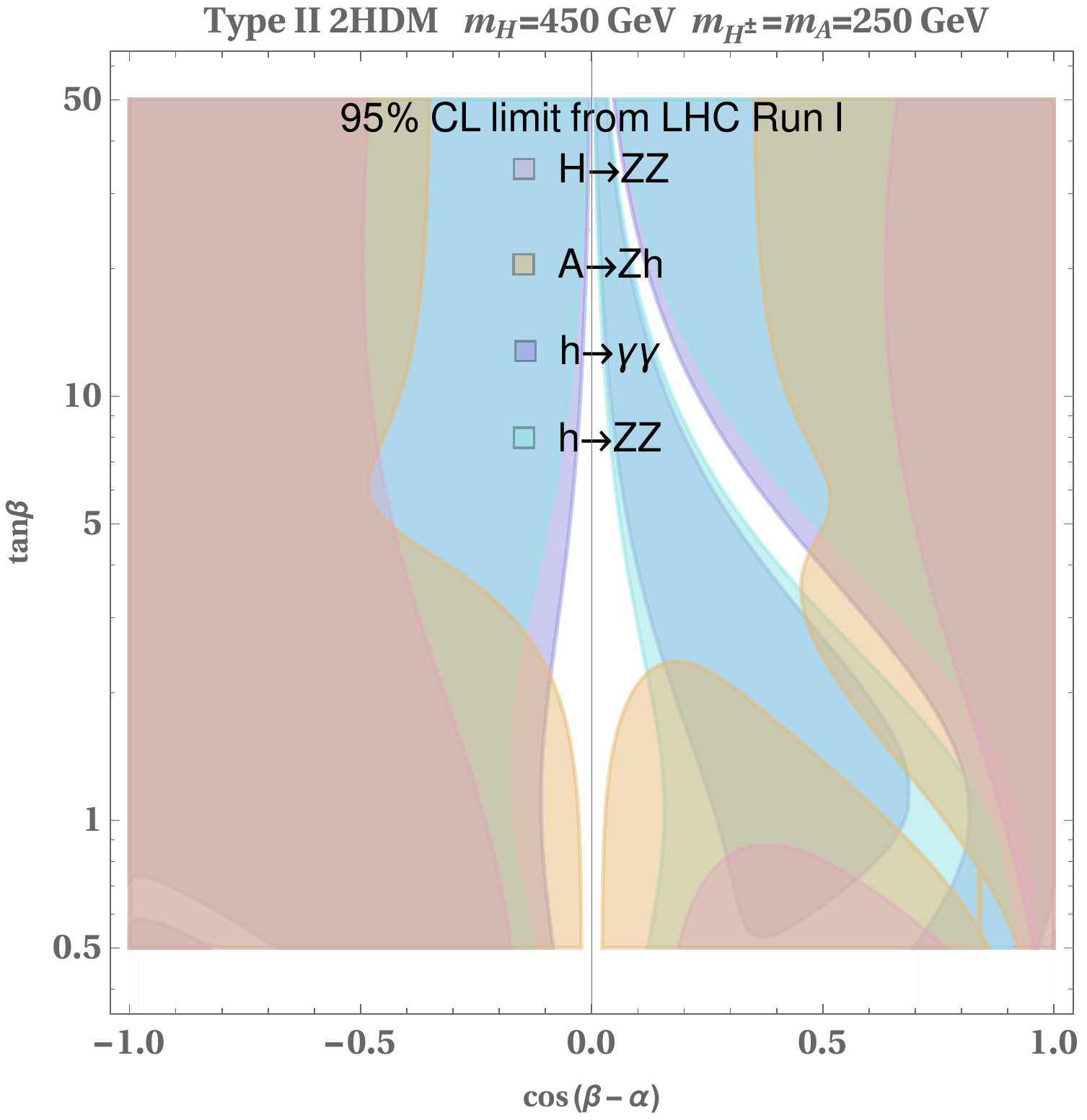}
\caption{Status of type II 2HDM after the 8~\TeV run of the LHC
for $m_A = m_{H^\pm} = 250~\GeV$, $m_H = 450~\GeV$.}
\label{type2}
\end{figure}

In type I models, the coupling of neutral Higgs fields to fermions are
\[
ffh \propto \frac{\cos\al}{\sin\be},
\qquad
ffH \propto \frac{\sin\al}{\sin\be},
\qquad
ffA \propto \cot\be,
\]
while in type II models the couplings depend on the type of fermion:
\[
\begin{split}
&uuh \propto \frac{\cos\al}{\sin\be},
\qquad
ddh,\ \ell\ell h \propto \frac{\sin\al}{\cos\be},
\\
& uuH \propto \frac{\sin\al}{\sin\be},
\qquad
ddH,\ \ell\ell H \propto \frac{\cos\al}{\cos\be},
\\
& uuA \propto \cot\be,
\qquad
ddA,\ \ell^+ \ell^-  A \propto \tan\be.
\end{split}
\]
The values of $\al$ and $\be$ determine much of the phenomenology, and
we will plot constraints in the plane of $\cos(\be - \al)$ and
$\tan\be$.  The vertical line $\cos(\be - \al) = 0$ is the alignment
limit, which is very difficult to probe experimentally.  In Figs.\ref{zzh}
and \ref{type2} we show the experimental constraints from the 8 TeV run of the
LHC on type I and II 2HDM models, for the benchmark masses $m_H =
450~\GeV$ and $m_{A} = m_{H^\pm} = 250~\GeV$.  The indirect
constraints from $h \to \ga\ga$ and $h \to ZZ$ are independent of the
masses of the heavy Higgs bosons, while the direct search constraints
from $H \to ZZ$ and $A \to Zh$ weaken rapidly when this mass is
increased.  Both direct and indirect probes are therefore essential
for probing the Higgs sector.

In type I models, we see from Fig.\ref{zzh} that there is a large region of
parameter space of the model at large $\tan\be$ that is currently
unconstrained experimentally.  The indirect constraints from $h \to
\ga\ga$ and $h \to ZZ$ are approximately independent of $\tan\be$ in
this region.  The constraint from $A \to Zh$ becomes ineffective at
large $\tan\be$ because the $Att$ coupling becomes small, suppressing
the dominant production gluon fusion production of $A$.  The
production of $H$ is unsuppressed, and the process $H \to ZZ$ is the
main direct constraints, but its effectiveness is limited by the large
background from SM $ZZ$ production.

Fig.\ref{type2} shows the same constraints for a type II model.  The indirect
constraints are much more constraining in this case, and there is much
less parameter space open for direct searches.  This is mainly due to
the fact that for large $\tan\be$ the $hb\bar{b}$ coupling becomes large
there, which affects the $h$ branching ratios.

In this paper we investigate the cascade decay $H \to ZA \to ZZh$ as a
way of probing the unconstrained region of the type I 2HDM.  The
motivation for this is that the $H$ production is unsuppressed at
large $\tan\be$.  The decay $A \to Zh$ is suppressed for
$\cos(\be - \al) \to 0$, but is still significant for much of the
unexplored region.  This can be seen in Fig.\ref{zzh_b}, which shows the $\si
\times \text{BR}$ for $gg \to H \to ZA \to ZZh$ superimposed on the other
experimental constraints.  We will demonstrate that this is more
effective than $gg \to H \to ZZ$ in this parameter range, because the latter
suffers from larger standard model backgrounds.

To compare the various channels, we use the benchmark mass values
$m_A = m_{H^\pm} = 250~\GeV$,
$m_H = 450~\GeV$.
We choose the other parameters in the 2HDM potential so that the decay $H \to
hh$ is negligible, which is easily accomplished without any large
couplings or tunings.  This benchmark model has a $\si \times
\text{BR}$ for $ZZh$ production of $0.45$~pb.  Additional properties of the
benchmark are given in Appendix A.  
We also give results from a scan over $m_H$ at the end of the paper.

\section{Searching for $ZZh$ Final States}

\noindent
To examine the experimental prospects for $ZZh$ production, we
calculate the expected signal significance for the benchmark model
with $m_A = 250~\GeV$ and $m_H = 450~\GeV$,
which has $\si \times {\rm BR}(ZZh) = 0.45~\rm{pb}$ for the
$14~\TeV$ LHC, with total luminosity of $300~\rm{fb}^{-1}$, as reported
in Table~\ref{zzh_finalstates}.  Simulated events for both signal and
SM backgrounds were generated by {\tt MadGraph5} \cite{madgraph}, with
showering and hadronization simulated by {\tt Pythia8} \cite{pythia},
and the detector response simulated by {\tt Delphes3} \cite{delphes}.
For the multi-jet backgrounds, jet matching was used.  The
tagging rate, particle reconstruction efficiency and particle
isolation requirement are reported by CMS in Run
I~\cite{cms3,cms4,cms5,cms6,cms7}.  The LO cross-sections of the SM
backgrounds for each channel are calculated by {\tt MadGraph5}.

To distinguish the Higgs Cascade signal over much larger SM
background, we apply event selection requirements, separately
optimized for each final state, as described below.

\subsection{Requirements on Reconstructed Particles}

Detector limitations, such as trigger requirements and detector
acceptance, are common across all channels, and so common selection
requirements are applied to reconstructed jets, muons, electrons, and
taus, before further selection requirements, optimized for each final
state, are applied.

Leptons are required to have $\pt > 10~\GeV$ and pseudorapidity
$|\eta|<2.5$. We further require isolated leptons, as determined from
the isolation ratio $R_{\rm iso} = p_{{\rm T},j}/p_{{\rm T},\ell}$
where $p_{{\rm T},j}$ is the clustered transverse energy, contained in
a cone of radius $\Delta R$ around the lepton, and $p_{{\rm T},\ell}$
is the lepton transverse energy.  The lepton isolation requirement
used in this analysis is $\Delta R<0.3$ with $R_{\rm iso}<0.1$, except
for signals where both $Z$'s decay to leptons. In this case, the
leptons are boosted in the $Z$ direction and therefore tend to be
close together.  We therefore use a different isolation requirement
with $\Delta R<0.2$, with $R_{\rm iso}<0.05$ for electrons and $R_{\rm iso}<0.1$ for
muons.  Similar isolation criteria have been used by ALTAS for their
multilepton searches in LHC Run I\cite{atlas6,atlas7}.

Jets are required to satisfy $\pt > 20~\GeV$ and $|\eta| < 5$. Photons
are required to satisfy $\pt > 10~\GeV$ and $|\eta| < 2.5$.  The
$b$-tagging rate is adjusted to be the same as one of the commonly
used $b$-taggings in CMS detector, which is $75\%$ tagging efficiency
and $1.5\%$ misidentification rate\cite{cms3}.  According to the $Z\to
\tau\tau$ cross section measurement in CMS run II, the hadronic
decayed tau identification efficiency is set to be $60\%$ and the
mis-identification rate from jets is set to be $3\%$.

The remaining event selection is optimized for each individual
channel, as described below.

\subsection{$ZZh \to (\ell^+ \ell^- )(\ell^+ \ell^- )(b\bar{b})$}

\noindent
This is the ``golden channel'' for the $ZZh$ final state, with a clean signal
and essentially no background. 
For this channel, we require
two $b$-tagged jets and two opposite-sign same-flavor (OSSF) isolated
lepton pairs.  The main backgrounds for this channel come from
$t\bar{t}Z$ and $ZZ+b\bar{b}$. As the probability of detecting a $b$
quark from misidentification of a jet is about $1.5\%$, the $ZZ + \text{jets}$
with two misidentified $b$ jets is much smaller than $ZZ+b\bar{b}$.
To ensure that the trigger efficiency is nearly $100\%$ for the
selected events, we require that the leading lepton has $\pt >
20~\GeV$, while the second has $\pt > 15~\GeV$.  

The remaining requirements were chosen to maximize the significance,
approximated by
\begin{equation*}
 Z = \sqrt{2 \left[(S+B)\times \ln\left(1+\frac{S}{B}\right)-S \right]}
\end{equation*}
where $S$ is the number of signal events and $B$ is the number of
background events, at $300~{\rm fb}^{-1}$. To suppress the $t\bar{t}Z$
background, the total missing transverse energy is required to satisfy
$\met < 80~\GeV$.

\begin{figure}[htb]
\centering
\begin{subfigure}[b]{0.47\textwidth}
\includegraphics[width=\textwidth]{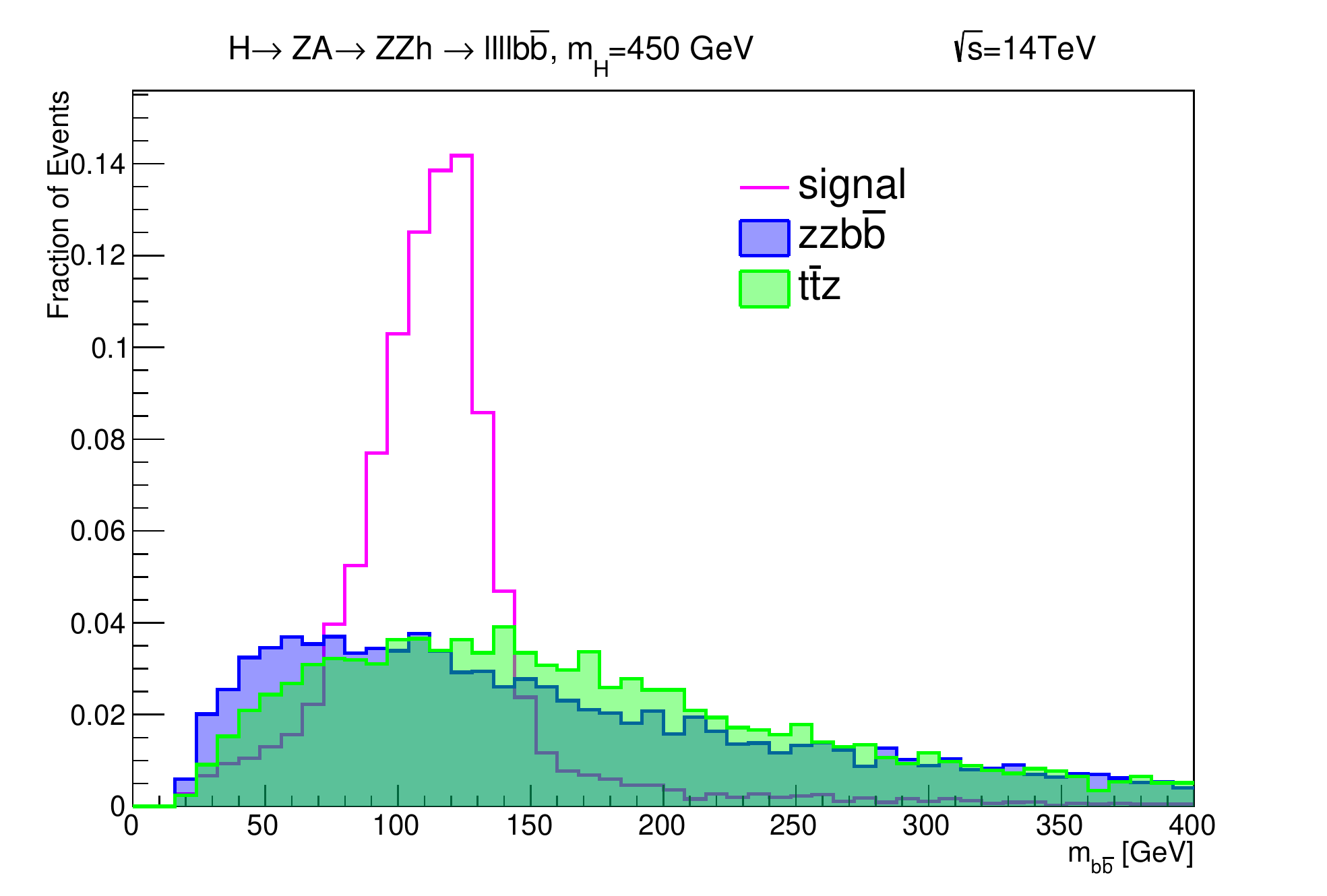}
\caption{Di-bottom invariant mass distributions. All normalized to 1.}
\label{dibotM}
\end{subfigure}
\begin{subfigure}[b]{0.47\textwidth}
\includegraphics[width=\textwidth]{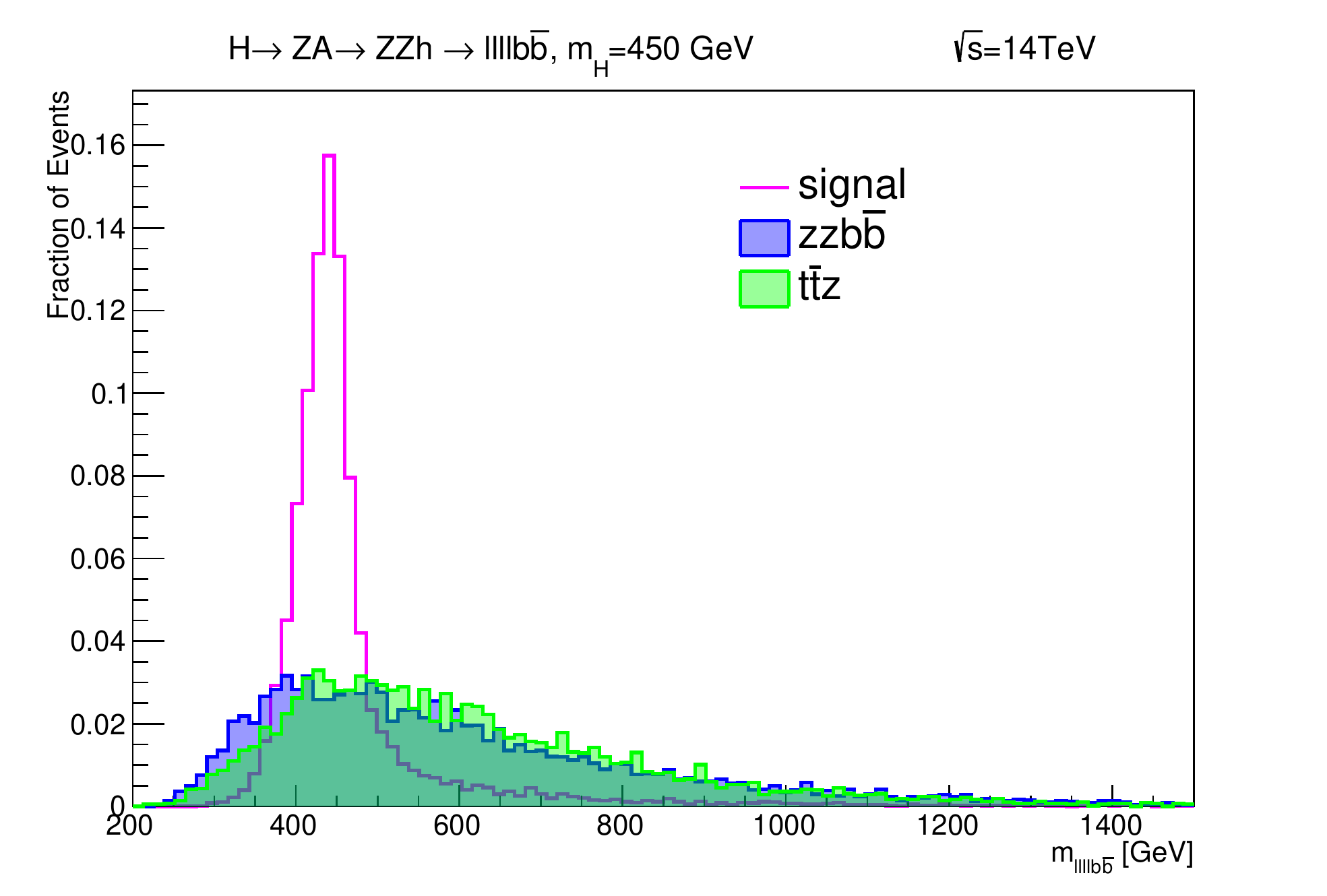}
\caption{Invariant mass distributions of $\ell^+ \ell^- \ell^+ \ell^-  b\bar{b}$. All normalized to 1.}
\label{mH}
\end{subfigure}
\caption{ \label{llllbb-kinematics}
Examples of kinematic variables studied for the decay channel $\ell^+ \ell^- \ell^+ \ell^-  b\bar{b}$. For the signal, both variables displayed here show prominent features that can be used to suppress the backgrounds.}
\end{figure}

In this final state, all of the intermediate particles in the Higgs
cascade are fully reconstructed.  This channel is therefore nearly
background free after cuts on the invariant mass of the reconstructed
$H$ and $A$.  The invariant mass of $b\bar{b}$ and $\ell^+ \ell^- \ell^+ \ell^- 
b\bar{b}$ for both signal and total background is presented in
Fig.~\ref{llllbb-kinematics}.  The optimized selection requirements
based on reconstructed invariant mass of the intermediate bosons are as follows:
\begin{compactitem} 
\item $m(b\bar{b}) \in [70, 160]~\GeV$
\item $m(\ell^+ \ell^- ) \in [71, 111]~\GeV$ for two $\ell^+ \ell^- $ pairs.
\item $m(\ell^+ \ell^-  b\bar{b}) \in [210,290]~\GeV$ for at least one $\ell^+ \ell^- $ pair.
\item $m(\ell^+ \ell^- \ell^+ \ell^-  b\bar{b}) \in [400,500]~\GeV$
\end{compactitem}

\begin{table}[htb]
\footnotesize 
\centering
\begin{adjustbox}{max width=\textwidth}
\begin{tabular}{|l|c|c|c|c|c|c|c|}
\hline\hline
$\#$ & 0& 1& 2& 3& 4& 5& 6\\
Selection& Initial& $\ell^+ \ell^- \ell^+ \ell^-  b\bar{b}$& $m_{b\bar{b}}$& $m_{\ell^+ \ell^- }$& $\met$&  $m_{A}$& $m_{H}$\\\hline
Signal& 353& 39.5& 33.6& 31.8& 29.5& 28.4& 23.9\\ \hline
$ttZ$& 643& 37.5& 14.3& 3.77& 1.51& 1.20& 0.53\\
$ZZb\bar{b}$& 81.6& 6.09& 2.18& 2.06& 1.70& 1.16& 0.43\\ 
Total background& 725& 43.6& 16.5& 5.83& 3.21& 2.35& 0.96\\
\hline\hline
\end{tabular}
\end{adjustbox}
\caption{Cut flows after each selection for the $\ell^+ \ell^- \ell^+ \ell^- 
  b\bar{b}$ channel. The number of signal and backgrounds are
  estimated for a luminosity of 300 $\mathrm{fb}^{-1}$. The third
  column gives the number of events after requiring 2 OSSF lepton
  pairs and 2 $b$-tagged jets.}
\label{llllbbcutflow}
\end{table}

Table~\ref{llllbbcutflow} displays the number of events after each cut
for both signal and backgrounds. The expected significance for our
benchmark model in this channel is $11\si$.  The integrated
luminosity required to reach $5\si$ is $66~\rm{fb}^{-1}$, and
$3\si$ is $24~\rm{fb}^{-1}$.

\subsection{$ZZh \to (\ell^+ \ell^- )(jj)(b\bar{b})$}

This channel benefits from the large branching ratios of $H \to
b\bar{b}$ and $Z \to jj$, but has significant $t\bar{t}jj$ and
$Zb\bar{b}jj$ background. Additional backgrounds such as
$W^+W^- + \text{jets}$ and $Z + \text{jets}$, where two jets are misidentified as
originating from a $b$-quark, were found to be negligible compared to
the two primary backgrounds.

We require two $b$-tagged jets, two non-$b$-tagged jets, and an OSSF
lepton pair.  To ensure that the trigger is nearly $100\%$ efficient,
the leading lepton must have $\pt>30~\GeV$.
 
\begin{figure}[H]
\centering
\begin{subfigure}[b]{0.47\textwidth}
\includegraphics[width=\textwidth]{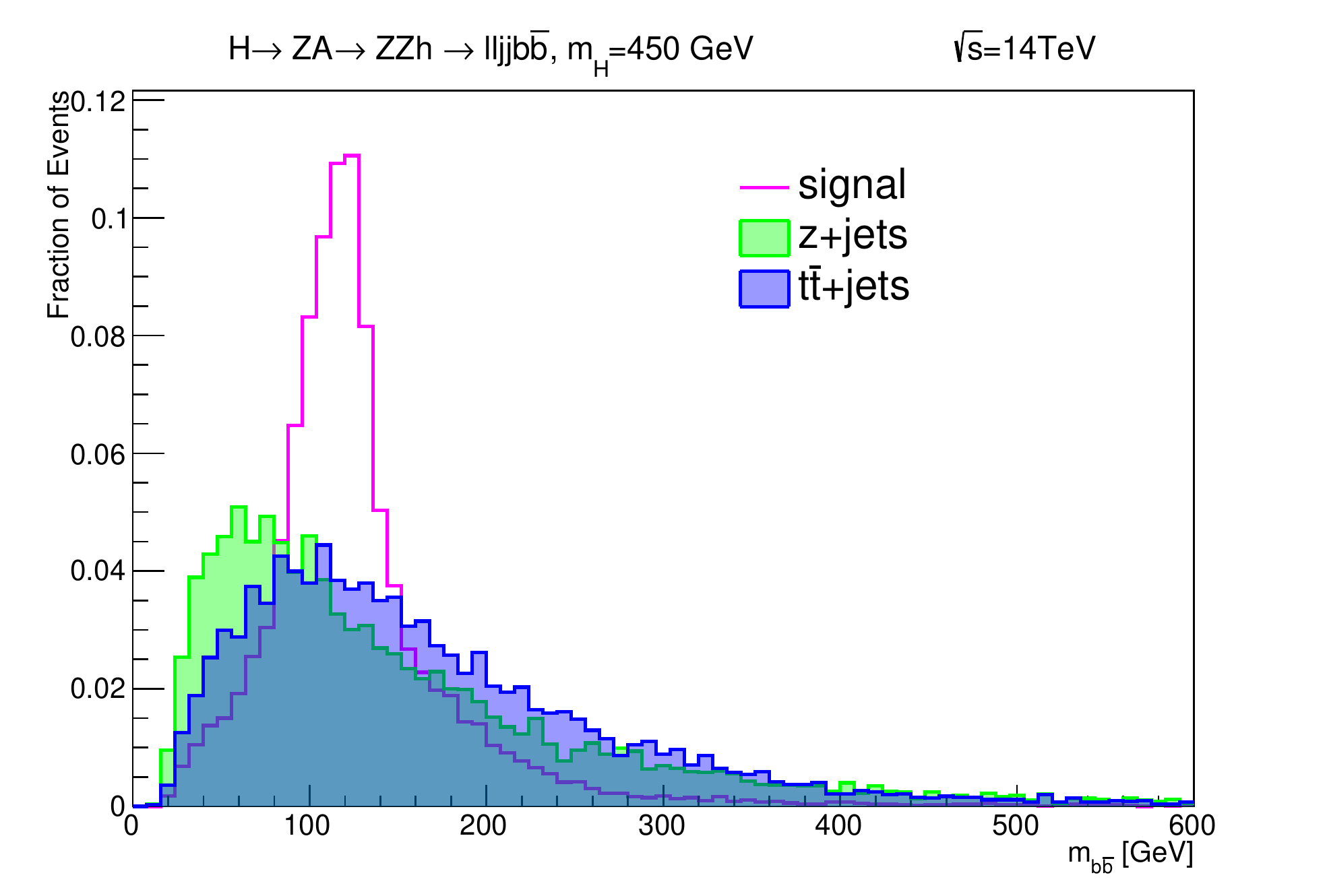}
\caption{Di-Bottom invariant mass distributions, normalized to one.}
\label{lljjbb1}
\end{subfigure}
\begin{subfigure}[b]{0.47\textwidth}
\includegraphics[width=\textwidth]{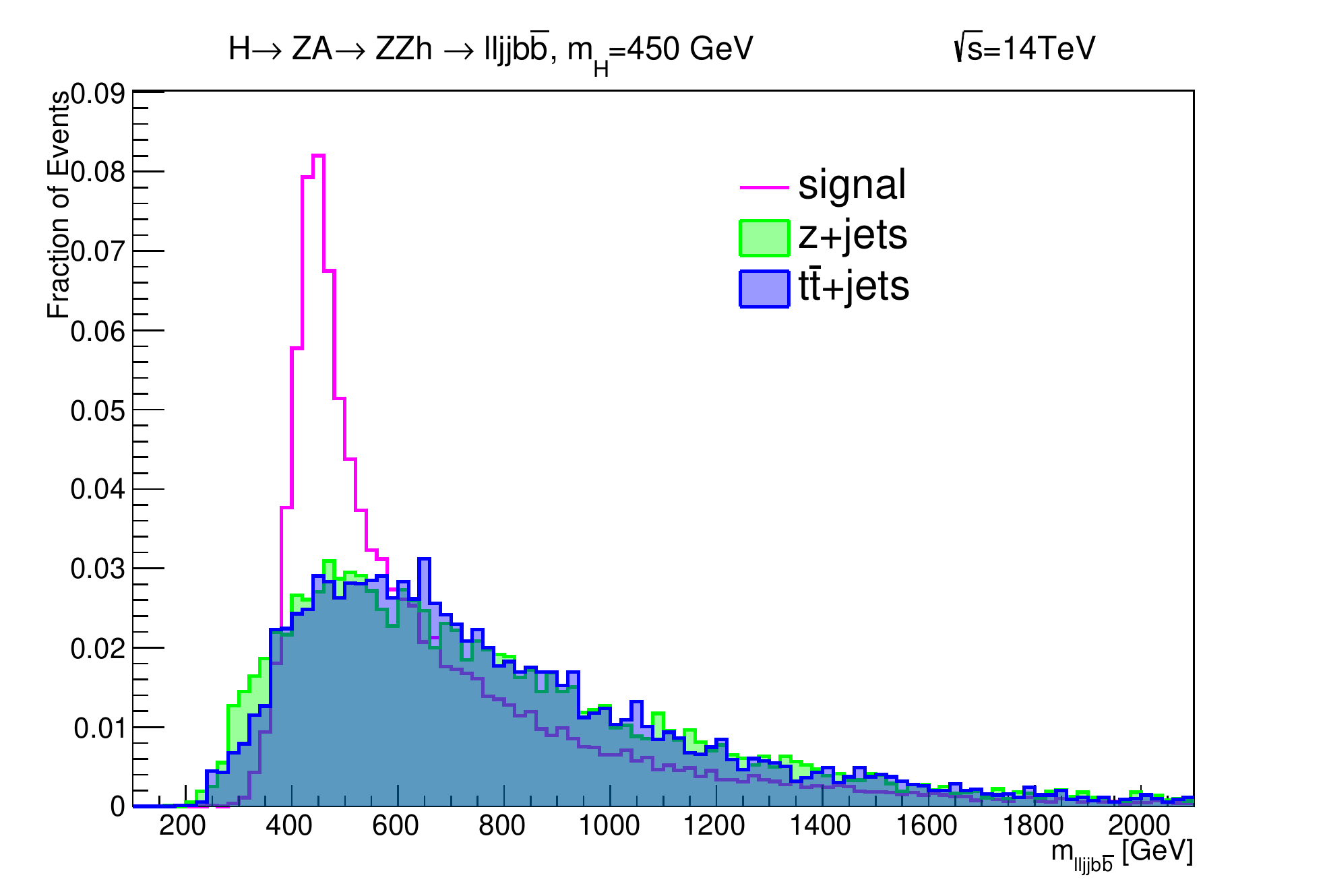}
\caption{$\ell^+ \ell^-  jjb\bar{b}$ invariant mass distributions, normalized to one.}
\label{lljjbb}
\end{subfigure}
\caption{Examples of kinematic variables studied for the decay channel
  $\ell^+ \ell^-  jj b\bar{b}$.}
\end{figure}

To devise selection requirements which optimally separate the Cascade
Higgs signal from SM background, several kinematic variables were
considered, including the $\pt$ of the $b\bar{b}$ system and its
invariant mass; the $\pt$ of the jet pair and its invariant mass; the
$\pt$ of the $\ell^+ \ell^- $ system and its invariant mass; the missing
transverse energy; and the reconstructed masses of $A$ and $H$.  Among
all the variables, $m_{b\bar{b}}$, $m_{\ell^+ \ell^- }$, $\met$, $m_A$ and
$m_H$ show the most promise for rejecting SM background.  We plot the
$m_{b\bar{b}}$ and $m_{\ell^+ \ell^-  jj b\bar{b}}$ distributions in Fig.~\ref{lljjbb1} and
Fig.~\ref{lljjbb}.

Because the background yields are much larger than the signal, the
approximation $S/\sqrt{B}$ is used to calculate the significance. The
optimized selection requirements based on reconstructed invariant mass and $\met$
are as follows:
\begin{compactitem} 
\item $m(b\bar{b}) \in [85, 160]~\GeV$
\item $m(\ell^+ \ell^- ) \in [81, 101]~\GeV$
\item $m(jj) > 60~\GeV $
\item $\met < 120~\GeV $
\item Either $m(\ell^+ \ell^-  b\bar{b}) < 390~\GeV$ or $m(jjb\bar{b}) < 390~\GeV$ 
\end{compactitem}
The event yields for each background and the signal are given in
Table~\ref{cutflowljb}.  The significance for this decay mode is
$3.6\si$ after the invariant mass selections on the candidate $A$
and candidate $H$. 

\begin{figure}[H]
\centering
\includegraphics[width=0.55\textwidth]{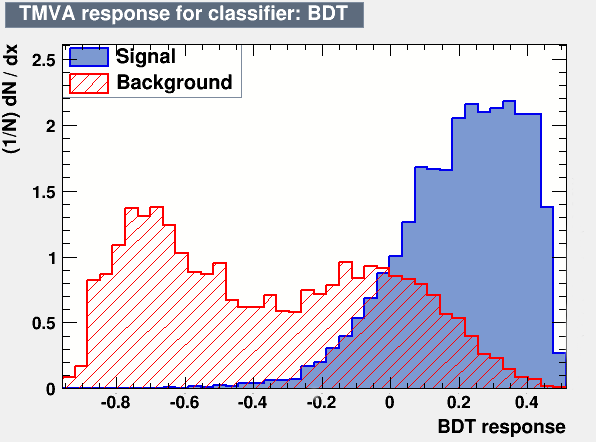}
\caption{The BDT response distribution with input variables of $m_{b\bar{b}}$, $m_{\ell^+ \ell^- }$, $m_{jj}$, $\met$, $m_A$ and
$m_H$.}
\label{bdt}
\end{figure}

In this channel, the background yield is much larger than the signal. So unlike the other channels, the uncertainty of the background may largely affect the outcome. In the calculation, the background uncertainty is set to be 5\% for cross section and 3\% for luminosity. By using likelihood ratio (LR) method, instead of simple $S/\sqrt{B} $, we calculate the significance to be $2.2\si$. Moreover, by using boosted decision tree (BDT) method, we find that the signal significance could reach $5.2\si$. In this method, the same five variables previously used in the cut-and-count analysis and $m_{\ell^+ \ell^- jj b\bar{b}}$ constitute the input of the BDT. The output, the BDT response, is used to calculate the significance with LR method, see Fig.~\ref{bdt}. 

\begin{table}[htb]
\footnotesize 
\centering
\begin{adjustbox}{max width=\textwidth}
\begin{tabular}{|l|c|c|c|c|c|c|c|}
\hline\hline
$\#$ & 0& 1& 2& 3& 4& 5& 6\\
Selection& Initial& $\ell^+ \ell^-  jjb\bar{b}$& $m_{b\bar{b}}$& $m_{\ell^+ \ell^- }$& $m_{jj}$& $\met$& $m_{A}$\\\hline
Signal & 5660& 916& 606& 560& 526& 510& 495\\ \hline
$Z+\text{jets}$& $7.1\times 10^6$& $8.8\times 10^4$& $2.3\times 10^4$& $2.2\times 10^4$& $1.8\times 10^4$&$1.7\times 10^4$ &$1.4\times 10^4$\\
$t\bar{t}jj$& $2.7\times 10^7$& $2.5\times 10^5$& $7.3\times 10^4$& $1.0\times 10^4$& $8.5\times 10^3$& $5.9\times 10^3$& $5.7\times 10^3$\\ \hline
Total background& $3.4\times 10^7$& $3.4\times 10^5$& $9.6\times 10^4$& $3.2\times 10^4$& $2.6\times 10^4$& $2.3\times 10^4$& $1.9\times 10^4$\\
\hline\hline
\end{tabular}
\end{adjustbox}
\caption{Cut flows after each selection for the $\ell^+ \ell^-  jj b\bar{b}$ channel. The number of signal and backgrounds are estimated for a luminosity of 300 $\mathrm{fb}^{-1}$. The third column gives the number of events after requiring an OSSF lepton pair, 2 $b$-tagged jets and 2 jets.}
\label{cutflowljb}
\end{table}

\subsection{$ZZh \to \ell^+ \ell^- \ell^+ \ell^-  + \met$}

In this channel, the relatively small
decay branching ratios involved are partially compensated by the
large number of combinations of decays which result in this final
state.  The main contribution is from cases where at least one $Z$
boson decays as $Z \to \ell^+ \ell^- $ while the Higgs decays either as $h
\to \ell \nu \ell \nu$ or $h \to \tau_\ell \tau_\ell$ or the other $Z\to \tau_\ell \tau_\ell$ where $\tau_\ell$
denotes a $\tau$ which subsequently decays leptonically.

The primary SM backgrounds are as follows::
\begin{compactitem} 
\item $ZZ$ production, with each $Z$ decaying either as $Z\to \ell^+ \ell^- $ or $Z \to \tau_\ell \tau_\ell$.
\item SM Higgs boson production with $h\to 4\ell$;
\item $Zh$ production with $Z\to \ell^+ \ell^- $ and $h\to \ell^+ \ell^- \nu\nu$ or $Z\to\nu\nu$ and $h\to 4\ell$;
\item $Wh$ production with $W\to l\nu$ and $h\to4\ell$.
\end{compactitem}

\begin{figure}[htb]
\centering
\includegraphics[width=0.8\textwidth]{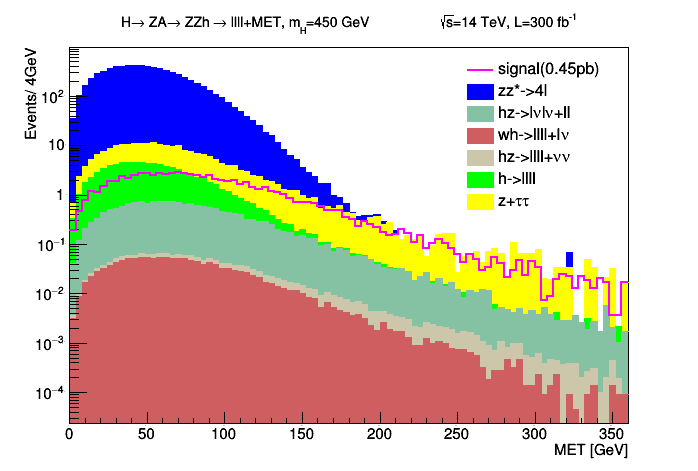}
\caption{$\met$ distribution for signal and backgrounds after 4 lepton selection with one OSSF at Z boson mass resonance. A high $\met$ cut is favored as can be seen. An integrated luminosity of 300 $\mathrm{fb}^{-1}$ is assumed.}
\label{llllm}
\end{figure}

The leading lepton must have $\pt > 20~\GeV$ and the sub-leading
lepton $\pt > 15~\GeV$.  In addition, at least one OSSF lepton pair is
required to have a reconstructed invariant mass in the range $81~\GeV
< m(\ell^+ \ell^- ) < 101~\GeV$.  The signal processes have larger $\met$
than most of the SM background, as shown in Fig.~\ref{llllm}, and the
selection requirement which maximizes the signal significance was
found to be $\met > 150~\GeV$.

Table~\ref{llllmet} lists the number of signal and background events
obtained after the final cuts. This significance of the benchmark
model after all cuts is $2.1\si$.

\begin{table}[htb]
\footnotesize 
\centering
\begin{adjustbox}{max width=\textwidth}
\begin{tabular}{|l|c|c|}
\hline\hline
 &Initial &  Final Yields  \\ \hline
Signal& 1600& 9.7\\\hline
$ZZ \to (\ell^+ \ell^-)(\ell^+ \ell^-)$ & $1.5\times 10^4$ & 8.2\\
$ZZ \to (\tau^+\tau^-)(\ell^+ \ell^-)$ & 7700 & 8.4\\
$h\to \ell^+ \ell^- \ell^+ \ell^- $& 240& 0.16\\
$Zh \to (\ell^+ \ell^-)(\ell^+ \ell^- \nu\nu)$ & 41& 1.3\\
$Zh \to (\nu\nu)(\ell^+ \ell^- \ell^+ \ell^-)$ & 1.3 &0.08\\
$Wh \to (\ell\nu)(\ell^+ \ell^- \nu\nu)$& 2.3& 0.09\\\hline
Total background& $2.3\times 10^4$ &18\\
\hline\hline
\end{tabular}
\end{adjustbox}
\caption{The event yields for the signal and backgrounds after the selection for the $\ell^+ \ell^- \ell^+ \ell^- + \met$ channel. 
An integrated luminosity of 300 $\mathrm{fb}^{-1}$ is assumed.}
\label{llllmet}
\end{table}

\subsection{$ZZh \to (\ell^+ \ell^- )(\nu\nu)(\ga\ga)$}
The main SM backgrounds for this channel are $t \bar{t} \ga j$ (where $j$ is misidentified as a photon), $t \bar{t} \ga \ga$, $W Z \ga \ga$, $\ell^+ \ell^-  \nu\bar{\nu}\ga\ga$ and $Z \ga \ga$. We also considered $\tau^+ \tau^- \ga\ga$ with the $\tau's$ decaying leptonically, but it was completely negligible after applying invariant mass cuts. For the $t \bar{t} \ga +\text{jets}$ background, we use $10^{-3}$ for the probability of a jet faking a photon, which is consistent with the CMS detector performance\cite{cms8}. For the $Z\ga \ga$  and the $\ell^+ \ell^- \nu\bar{\nu}\ga\ga$ backgrounds, we also take into account the contribution from $ h\to \ga\ga$. 

Despite of a very clean signature, this channel suffers from very small cross sections for both the signal and backgrounds. We have applied $\pt$ cuts for the photons and leptons, invariant mass cuts for the two leading charged leptons and photons, and a MET cut to take advantage of the neutrinos in the final state. After optimizing the cuts for the significance, the final selections are as follows:
\begin{compactitem}
  \item Two photons with $\pt>17\mbox{ \GeV}$
  \item Two leptons with $\pt>13\mbox{ \GeV}$
  \item Missing $E_T>50\mbox{ \GeV}$
  \item Invariant mass $m_{\ga\ga} \in [120, 130]$~\GeV
  \item Invariant mass $m_{\ell^+ \ell^- } \in [84, 98]$~\GeV
\end{compactitem}

Table~\ref{tab:nnllaa} lists the number of events obtained after the final selection for the signal and the backgrounds. Despite of a sharp drop in the signal cross section due to the tiny $h\to \ga\ga$ branching ratio ($\sim10^{-3}$), the good resolution in the reconstructed $M_{\ga\ga}$ is especially effective in suppressing the backgrounds. The significance of this channel was found to be $2.8\si$, 
provided that the signal yields is 2.1 and that the background yields is 0.14.

\begin{table}[H]
\footnotesize 
\centering
\begin{adjustbox}{max width=\textwidth}
\begin{tabular}{|l|c|c|c|c|c|c|}
\hline\hline
$\#$		& 0			& 1 & 2 & 3 & 4 & 5\\
Selection 	& Initial & $\ell^+ \ell^- \ga\ga $ & $p_{T_{\ell,\ga}}$ & $m_{\ga\ga}$ & $m_{\ell^+ \ell^- }$ & $\met$ \\
\hline
Signal			 & 8.3	& 3.3 & 3.1 & 2.8 & 2.5 & 2.1 \\\hline
$t \bar{t} \ga +\text{jets}$ (1 fake $\ga$)	 & 41 & 10 & 6.8 & 0.27 & 0.078 & 0.059 \\

$Z \ga \ga$ & 106 & 56 & 46 & 20 & 18 & 0.028\\

$t \bar{t} \ga \ga$  & 2.5 & 0.51 & 0.39 & 0.13 & 0.038 & 0.028 \\

$\ell^+ \ell^-  \nu\bar{\nu}\ga\ga$    & 0.48 & 0.25 & 0.19 & 0.078 & 0.035 & 0.024\\

$W Z \ga \ga$     & 0.064 & 0.041 & 0.030 & $9.1\times 10^{-3}$ & $5.1\times 10^{-3}$ & $2.4\times 10^{-3}$ \\
\hline
Total background   & 150  & 67 & 43 & 20 & 18 & 0.14 \\
\hline \hline
\end{tabular}
\end{adjustbox}
\caption{Cut flows for the $\ga\ga\nu\nu\ell^+ \ell^- $ channel. An integrated luminosity of $300\mbox{ fb}^{-1}$ is assumed. The third column gives the number of events obtained after requiring an OSSF lepton pair and two photons.}
\label{tab:nnllaa}
\end{table}

\subsection{$ZZh \to (\ell^+ \ell^- )(jj)(\ga\ga)$}
This channel faces the similar challenges as the $\ell^+ \ell^- \nu\nu\ga\ga$ channel, where both the signal and the background cross sections are fairly small. However, the sharp resolution in $M_{inv}(\ga\ga)$ for the signal helps significantly suppress the backgrounds. The main SM backgrounds for this channel contain the irreducible background $\ell^+ \ell^-  j j \ga\ga$, $\ell^+ \ell^-  \ga+\text{jets}$ with one jet-faked photon and $\ell^+ \ell^- +\text{jets}$ with two fake photons. Other SM processes, such as $t\bar{t}+$ fake $\ga$s, $t\bar{t}\ga\ga$ and $ZZ\ga+$ fake $\ga$ can also contribute but their final yields are much smaller compared to the main backgrounds due to their smaller cross sections and the tiny jet-faking-photon rate. Similarly to the $Z\to jj$ decay in the $\ell^+ \ell^-  jj b\bar{b}$ final state search, the reconstructed $m_{jj}$ has so broad a peak that it cannot be exploited. The final selections are as follows:
\begin{compactitem}
  \item Two photons with $\pt>24\mbox{ \GeV}$
  \item Two jets with $\pt>24\mbox{ \GeV}$
  \item Two leptons with $\pt>13\mbox{ \GeV}$
  \item Invariant mass $m_{\ga\ga} \in [120, 130]$ \GeV
  \item Invariant mass $m_{\ell^+ \ell^- } \in [81, 101]$ \GeV
  \item Missing $E_T<120\mbox{ \GeV}$
\end{compactitem}
Table~\ref{tab:jjllaa} shows the cut flows and signal and background yields for this channel. With $5.7$ signal events and $9.6$ background events, we get a $1.7\ \si$ significance.

\begin{table}[H]
\footnotesize 
\centering
\begin{adjustbox}{max width=\textwidth}
\begin{tabular}{|l |c| c| c| c|c| c|}
\hline\hline
$\# $& 0&1&2&3&4&5\\
 Selection& Initial & $\ga\ga j j \ell^+ \ell^- $ & $p_{T_{\ell,\ga,j}}$& $m_{\ga\ga}$ & $m_{\ell^+ \ell^- }$ & $\met$ \\
\hline
Signal	& 29 & 8.3 & 6.8 & 6.2 & 5.7 & 5.7 \\\hline
$\ell^+ \ell^-  jj \ga\ga$   & 230 & 38 & 18 & 9.4 & 7.1 & 7.1\\
$\ell^+ \ell^- \ga jjj$ (1 fake $\ga$) &12900 & 267 & 66 & 2.6 & 1.9 & 1.9\\
$\ell^+ \ell^-  + \text{jets}$ (2 fake $\ga$s) & 4640 & 24 & 14 & 0.53 & 0.48 & 0.48\\
$t\bar t \ga j$ (1 fake $\ga$) & 25 & 4.0 & 2.5 & 0.11 & 0.054 & 0.043\\
$t\bar t \ga\ga$ & 1.1 & 0.20 & 0.13 & 0.062 & 0.031 & 0.024\\
$t\bar t + \text{jets}$ (2 fake $\ga$s)   & 19 & 2.1 & 1.6 & 0.055 & 0.027 & 0.022\\
$ZZ\ga j$ (1 fake $\ga$) & 1.4 & 0.52 & 0.32 & 0.013 & 0.012 & 0.012\\
\hline
Total background & 17800 & 335 & 103 & 13 & 9.6 &  9.6\\
\hline \hline
\end{tabular}
\end{adjustbox}
\caption{Cut flows for the $\ga\ga jj \ell^+ \ell^- $ channel. An integrated luminosity of $300\mathrm{fb}^{-1}$ is assumed.}
\label{tab:jjllaa}
\end{table}

\subsection{Other Channels}
There exist other channels that do not perform as well as the previously described 5 channels, nevertheless show interesting features that are worth considering given the prospect of a higher integrated luminosity.

 One of them is $Z\to \nu\nu,\,Z\to jj,\,h\to\ga\ga$. As in the case of $\ell^+ \ell^-  j j b \bar{b}$ (Fig.~\ref{lljjbb}), the reconstructed $Z$ from dijet is almost useless in differentiating between the signal and the backgrounds. Therefore, this channel suffers not only from a small cross section, but also from ``wasting" two $Z$s. However, just as those channels with $h\to \ga\ga$, the outstanding resolution in the reconstructed $m_h$ can be very powerful in suppressing the background. The event selection requires two photons that satisfy $122 < m_{\ga\ga} < 128$ \GeV, two jets that satisfy $m_{jj} > 60$ \GeV and a $\met$ requirement of $\met > 140$ \GeV. Due to the massive cross section, QCD production of two photons still dominates after the selection of $m_h$ and high $\met$ requirement. The final signal yield is 7.4 while background yield is $240$, giving a significance of $0.5\si$ assuming an integrated luminosity of 300 fb$^{-1}$. Table~\ref{tab:nnjjaa} displays the cut flows for the signal and the total background for this channel.
 
\begin{table}[H]
\footnotesize 
\centering
\begin{adjustbox}{max width=\textwidth}
\begin{tabular}{|l |c| c| c| c|c| c|c|}
\hline\hline
$\# $& 0&1&2&3&4\\
 Selection& Initial & $jj\ga\ga$& $m_{\ga\ga}$ & $m_{jj}$ & $\met$ \\
\hline
Signal& 67.8& 38.9& 28.6& 27.4& 7.4  \\\hline
Total background& $3.8\times 10^{12}$& $2.6\times 10^7$& $3.3\times 10^5$  & $2.9\times 10^5$& $2.4\times 10^2$ \\
\hline \hline
\end{tabular}
\end{adjustbox}
\caption{Cut flows for the $\ga\ga \nu\nu jj$ channel. An integrated luminosity of $300\mathrm{fb}^{-1}$ is assumed.}
\label{tab:nnjjaa}
\end{table}

 Another interesting decay mode is $Z\to \ell^+ \ell^- ,\,Z\to jj,\,h\to\tau_h\tau_h$. This channel utilizes the hadronic tau tagging in LHC detectors. Compared to $Z\to \ell^+ \ell^- ,\,Z\to jj,\,h\to b\bar{b}$, it successfully avoids the large background from $t\bar{t}+\text{jets}$. But it is seriously inflicted by the high tau mis-tagging rate. Hence, $Z+\text{jets}$ with two misidentified $\tau_h$s dominates. The event selection requires an OSSF lepton pair that satisfies $76 < m_{\ell^+ \ell^- } < 106$ \GeV, an OSSF tau pair that satisfies $80 < m_{\tau_h\tau_h} < 150$ \GeV, two non tau-tagged jets, $\met < 130$ \GeV, $m_A < 330$ \GeV and $m_H < 670$ \GeV. $m_{\ell^+ \ell^- }$ and $\met$ selection is applied to ensure that the other SM backgrounds are negligible compared to DY lepton pair production. The final significance is about $0.7\si$.  Table~\ref{tab:lljjtt} gives the the cut flows for the signal and the backgrounds.
 
\begin{table}[H]
\footnotesize 
\centering
\begin{adjustbox}{max width=\textwidth}
\begin{tabular}{|l |c| c| c| c|c| c|c|}
\hline\hline
$\# $& 0&1&2&3&4&5&6\\
 Selection& Initial & $\ell^+ \ell^-  jj\tau\tau$ & $m_{\tau\tau}$ &$m_{\ell^+ \ell^- }$ &  $\met$ & $m_A$& $m_H$\\
\hline
Signal& 630	&26& 18& 17& 17& 15& 11  \\\hline
Total background& $2.3\times 10^6$& $4.9\times 10^3$& $1.3\times 10^3$& $1.3\times 10^3$& $1.2\times 10^3$& 620& 290 \\
\hline \hline
\end{tabular}
\end{adjustbox}
\caption{Cut flows for the $\ell^+ \ell^-  jj\tau_h\tau_h$ channel. An integrated luminosity of $300\mathrm{fb}^{-1}$ is assumed.}
\label{tab:lljjtt}
\end{table}

\subsection{Results of the Benchmark Study}
In Table~\ref{zzh_results}, we list the number of signal and background yields for each channel studied and the significance reached at both 300 $\mathrm{fb}^{-1}$ and 3000 $\mathrm{fb}^{-1}$.

\begin{table}[H]
\footnotesize 
\centering
\begin{adjustbox}{max width=\textwidth}
\begin{tabular}{|l |c c c|c c c|}
\hline\hline
Channel &  & 14 TeV 300$\mathrm{fb}^{-1}$ & & & 14 \TeV\  3000$\mathrm{fb}^{-1}$ & \\ \hline
& Sig Yields & Bkg Yields & Significance & Sig Yields & Bkg Yields & Significance\\ \hline
$\ell^+ \ell^- \ell^+ \ell^-  b\bar{b}$ & 24 & 0.96 &  11 $\si$& 240 & 9.6 &  34 $\si$ \\ \hline
$\ell^+ \ell^-  jj b\bar{b}$ & 495 & 1.9$\times 10^4$ &3.6 $\si$ &4950 & 1.9$\times 10^5$ &11 $\si$ \\ \hline
$\ell^+ \ell^- \ell^+ \ell^- +\met$ & 9.7 & 18 &2.1 $\si$ & 97 & 180 & 6.6 $\si$ \\ \hline
$\ell^+ \ell^- \nu\nu\ga\ga$ & 2.1 & 0.14 & 2.8 $\si$  & 21  & 1.4 & 9.1 $\si$ \\ \hline
$\ell^+ \ell^-  jj \ga\ga$ & 5.7 & 9.6 & 1.7 $\si$ & 57 & 96 & 5.3 $\si$\\
\hline\hline
\end{tabular}
\end{adjustbox}
\caption{ Summary of the most sensitive channels of the $H\to ZA$, $A \to Zh$ cascade search for a benchmark $m_H = 450$ \GeV, $m_A=m_{H^{\pm}}=250$ \GeV, assuming the $ZZh$ production cross section is $0.45$ pb for the signal.}
\label{zzh_results}
\end{table}

 Among all the channels studied, $\ell^+ \ell^- \ell^+ \ell^-  b\bar{b}$ performs the best; the four-lepton requirement provides a clean signal, and $h\to b\bar{b}$ has a relatively large branching ratio. The next best performing channel is $\ell^+ \ell^-  jj b\bar{b}$. Although $\ell^+ \ell^-  jj b\bar{b}$ gives the most signal events, it is hard to suppress the $Z+\text{jets}$ backgrounds, for the jet shower is severe, the resolution of jet energy is not as good as that of leptons, and the reconstructed dijet mass peak is not significant enough to suppress the backgrounds. $\ell^+ \ell^- \nu\nu\ga\ga$ and $\ell^+ \ell^-  jj \ga\ga$ enjoy an outstanding resolution in the reconstructed $m_h$, meanwhile suffering from a much smaller cross section compared to the three other channels. 
 
 For the most sensitive channel $\ell^+ \ell^- \ell^+ \ell^-  b\bar{b}$, we compare the significance reach of this channel with those of the direct $A$ and $H$ searches at 14~TeV for our benchmark. For simplicity, we only consider $A\to Zh\to \ell^+ \ell^-  b\bar{b}$ and $H\to ZZ\to \ell^+ \ell^- \ell^+ \ell^- $ for the direct searches, both of which were studied by CMS and ATLAS in LHC Run I. Details of the direct searches considered here can be found in Appendix B and C. 
 In Fig.~\ref{expbounds_14},  we give projections of the $A$ and $H$ direct searches, assuming an integrated luminosity of 300 $\mathrm{fb}^{-1}$. We also display the ATLAS' projections of 95\% CL likelihood contours for $\kappa_V$ and $\kappa_F$\footnote {$\kappa_F$ ($\kappa_V$) is the ratio between the Higgs to fermions (vector bosons) coupling to its SM value.} at 14 TeV with 300~$\mathrm{fb}^{-1}$\cite{atlas4} 
 in the same plot. 
In Fig.~\ref{llllbb}, the limits that can be reached by the $H$ cascade search are shown side by side with the direct searches. It can be clearly seen that the $\ell^+ \ell^- \ell^+ \ell^-  b\bar{b}$ channel from the $H$ cascade search is more sensitive for $\tan\be \gtrsim 8$.

\begin{figure}[H]
\centering
\begin{subfigure}[b]{0.45\textwidth}
\includegraphics[width=\textwidth]{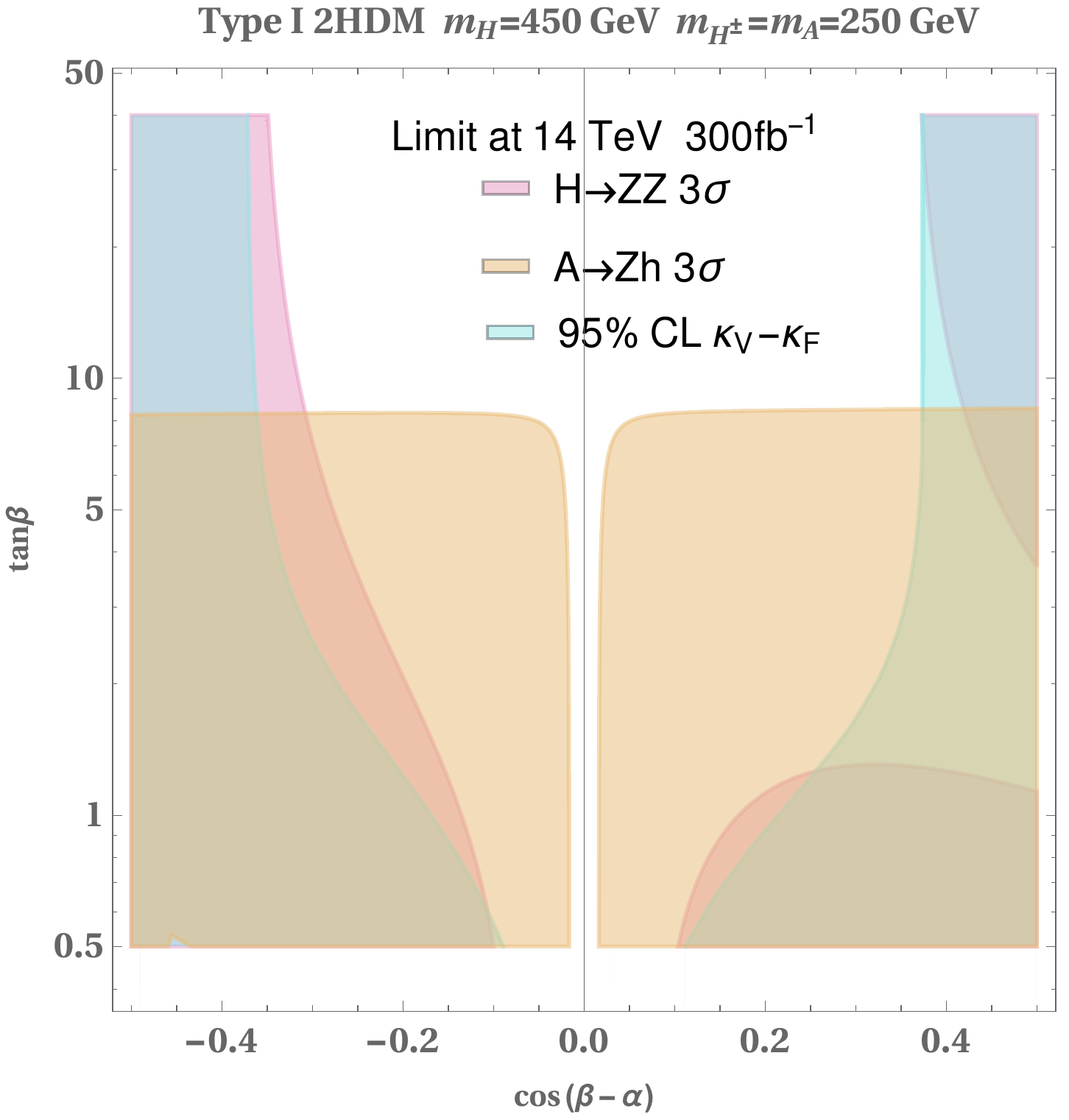}
\caption{}
\label{expbounds_14}
\end{subfigure}
\begin{subfigure}[b]{0.45\textwidth}
\includegraphics[width=\textwidth]{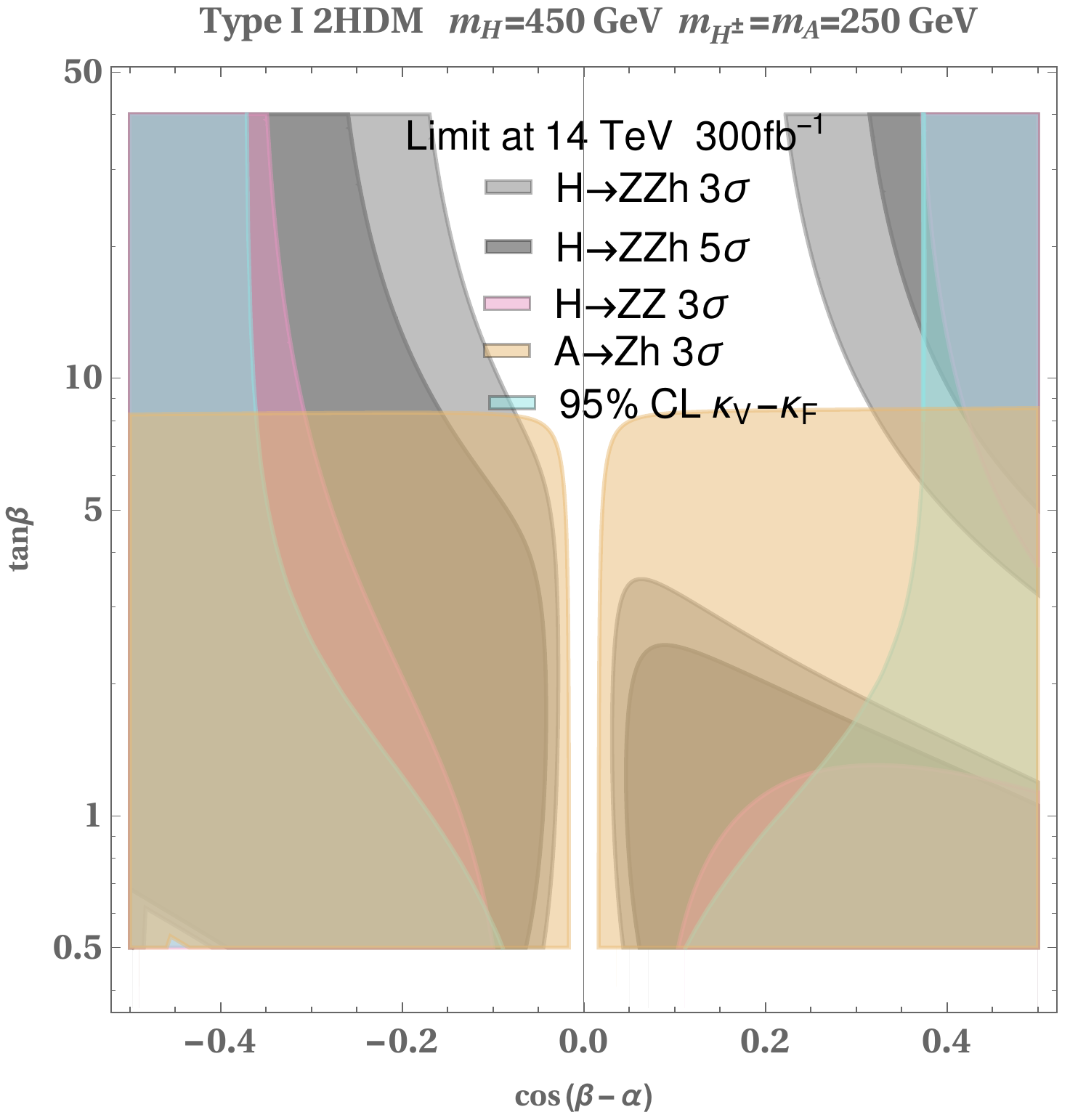}
\caption{}
\label{llllbb}
\end{subfigure}
\caption{The projections of the $A$ and $H$ direct searches and the SM Higgs coupling measurements at LHC run II, for $m_H=450$ \GeV, $m_A=m_{H^{\pm}}=250$ \GeV. Plotted on the right are the 3 and 5 $\si$ limits reached by the $H$ cascade search. }
\end{figure}

\begin{figure}[H]
\centering
\includegraphics[width=0.45\textwidth]{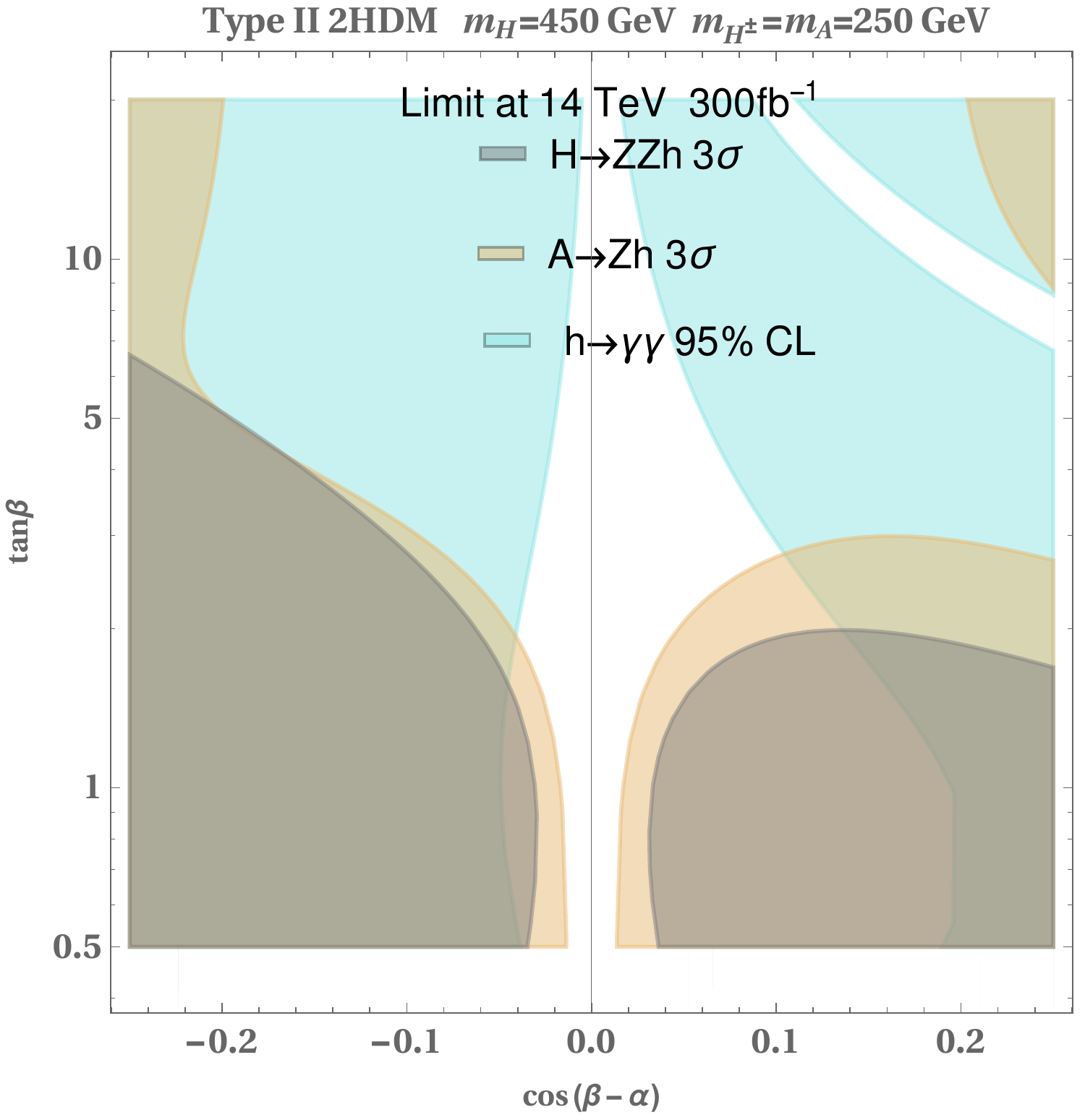}
\caption{The projections of the direct $A$ search, the $H$ cascade search and the ATLAS's $h\to \ga\ga$ coupling measurement projected at LHC run II, for $m_H=450$ \GeV, $m_A=m_{H^{\pm}}=250$ GeV.}
\label{llllbb_2}
\end{figure}

To demonstrate that our search is less sensitive in typeII model, we compare the significance reach of the $\ell^+ \ell^- \ell^+ \ell^-  b\bar{b}$ channel from the $H$ cascade search with the direct search $A\to Zh\to \ell^+ \ell^-  b\bar{b}$ for our benchmark in Fig.~\ref{llllbb_2}. We also display the ATLAS' projection on $h\to \ga\ga$ cross section at 14 TeV in the same plot. As one can see, the direct $A$ search is slightly more sensitive compared to the $H$ cascade search.

\begin{figure}[H]
\centering
\includegraphics[width=0.45\textwidth]{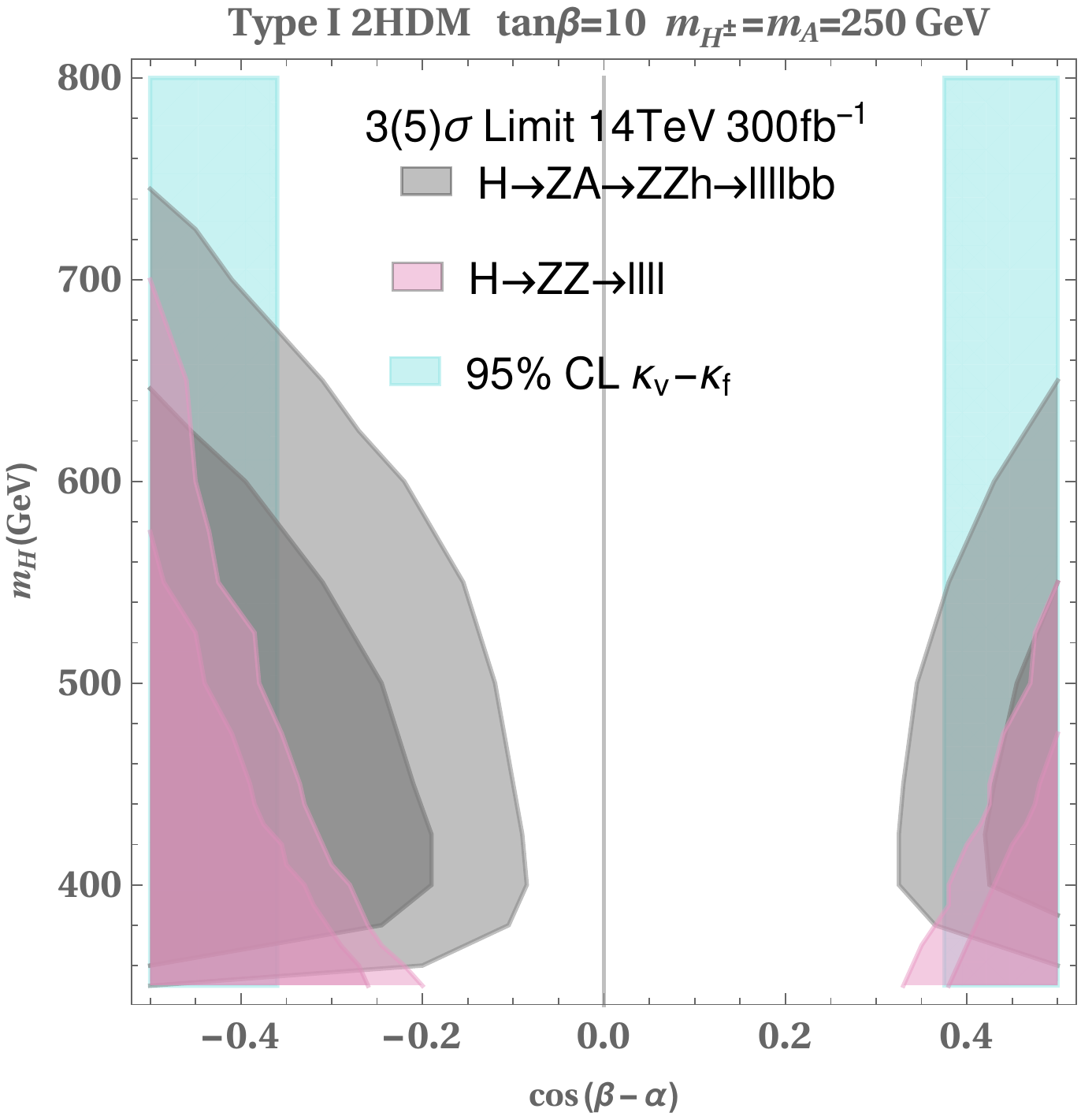}
\captionof{figure}{The limits that can be reached by the $H$ cascade search as a function of its mass with fixed $\tan\be$, $m_A$ and $m_{H^{\pm}}$, assuming an integrated luminosity of 300 $\mathrm{fb}^{-1}$. For comparison, the limits are also plotted for the direct $H$ search and the ATLAS's projections on SM Higgs coupling measurements.}
\label{mH_cos}
\end{figure}

The benchmark study shows that the $H$ cascade search is most
sensitive at large $\tan\be$.  We consider as a benchmark
$\tan\be=10$, and investigate the range in $m_H$ for which this search
provides new sensitivity.  For $\tan\be=10$, the direct $A$ search is
completely ineffective (see Fig.~\ref{llllbb}) because the production
of $A$ via quark loops is suppressed by $\tan^2\be $ in Type I
2HDM. That leaves the direct $H$ search as the only relevant
comparison to the $H$ cascade search for this benchmark.  As can be
seen from Fig.~\ref{mH_cos}, the 3 $\si$ exclusion of the $H$ cascade
search reaches $m_H\sim 650$ \GeV within two standard deviations of
the SM limit. The direct $H$ search is not as sensitive as the cascade
search, as it suffers from the SM $ZZ$ background.

The theory makes no prediction for the value of $m_H$, and therefore
the look-elsewhere effect should be considered.  However, considering
the optimized selection criteria for $m_{\ell^+ \ell^- \ell^+ \ell^-
  b\bar{b}}$ in the most sensitive channel, we estimate that the
number of independent $m_H$ bins in the region this analysis is
sensitive is only about three.  Thus, the look-elsewhere effect is not
expected to significantly affect the projected sensitivity of this
search.

\section{Conclusions}
In this paper, we demonstrate that the search $gg\to H\to AZ\to ZZh$ can be a discovery mode for 
additional Higgs boson at the 14~TeV LHC. 
For a benchmark with $m_A = 250~\GeV$ and $m_H = 450~\GeV$ we find that the significance with
300~fb$^{-1}$ is over $10\sigma$ in the ``golden channel'' $(\ell^+\ell^-)(\ell^+\ell^-)(b\bar{b})$, and that the most signal abundant channel $(\ell^+\ell^-)(jj)(b\bar{b})$ may reach $5\si$ significance with a boosted decision tree analysis.
At the high-luminosity LHC (3000~fb$^{-1}$) the prospects are even better,
with five channels having over $5\si$ significance.
We emphasize that this benchmark will not be probed by other search channels or Higgs coupling
measurements.
The reach also persists for values of $m_H$ up to $650~\GeV$, again in models that are not
probed by existing searches.
We conclude that searches for $ZZh$ final state are highly motivated at the 14~TeV
LHC.

\section*{Acknowledgements}
We are grateful to E. Salvioni and Y. Tsai for their help at early
stages of this work.
M.A.L. wishes to thank the Galileo Galilei Institute for hospitality.
This work was supported by DOE
grant DE-FG02-91ER40674.

\clearpage

\begin{figure}[H]
\centering
\begin{subfigure}[h]{0.45\textwidth}
\includegraphics[width=\textwidth]{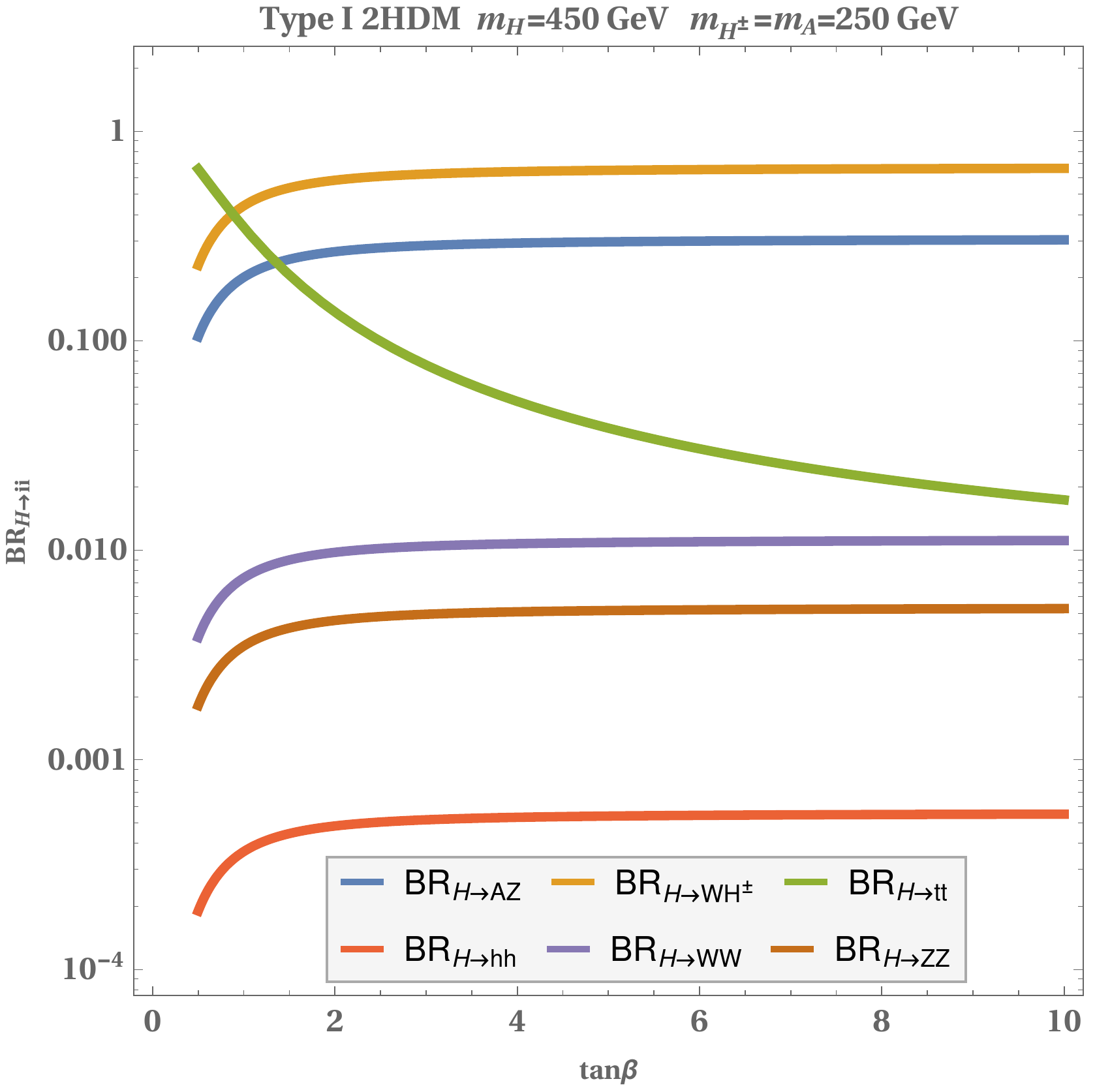}
\caption{$cos(\be-\alpha)=-0.1$.}
\label{heavyHbr1}
\end{subfigure}
\begin{subfigure}[h]{0.45\textwidth}
\includegraphics[width=\textwidth]{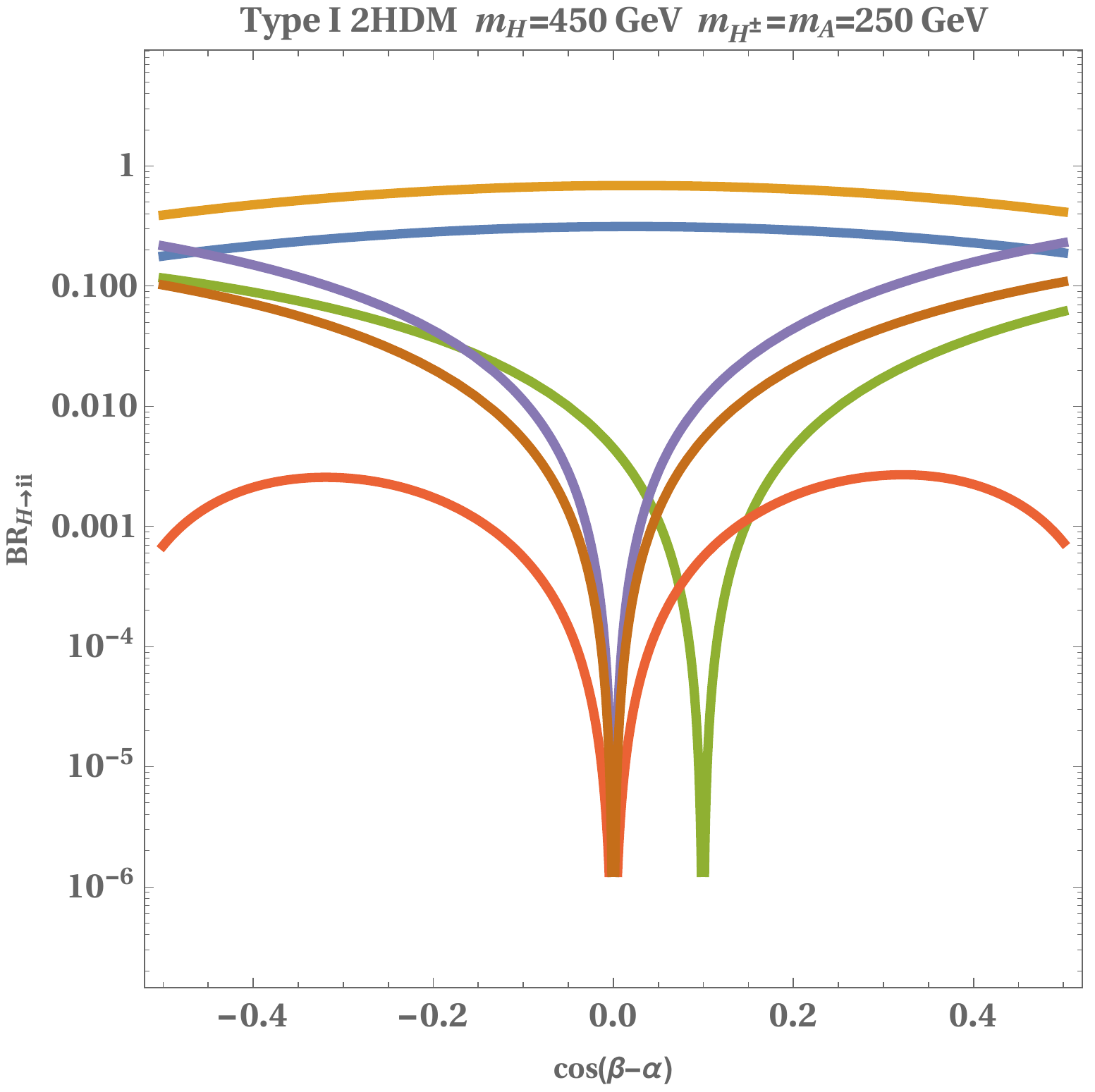}
\caption{$tan\be =10$.}
\label{heavyHbr2}
\end{subfigure}
\caption{Branching Ratios of $H$.}
\label{hbr}
\end{figure}

\appendix{Appendix A: Properties of the benchmark}
Four our study we choose a benchmark model where the masses
of all physical Higgs bosons are fixed. 
We choose $m_A = 250$~GeV so that the decay $A\to Zh$ is open
but the decay $A \to t\bar{t}$ is closed.
We choose $m_H = 450$~GeV so that the decay $H \to ZA$ is open.
These masses are chosen near the lower end of the lowest range for which the cascade decay
$H \to ZA \to ZZh$ is allowed in order to maximize the rate.
The main constraint from precision electroweak data comes from custodial symmetry breaking
\cite{haber10}, and this is suppressed by choosing $m_{H^{\pm}} = m_A$. 
With these masses fixed, there is still sufficient freedom to suppress the Higgs self-couplings, 
so that the $H\to hh$ decay is unimportant.
To illustrate this, in Figure~\ref{hbr}, the $H$ branching ratios are plotted at a fixed 
$\cos(\be-\al)$ and $\tan\be$, respectively.

\appendix{Appendix B: $A\to Zh\to \ell^+ \ell^-  b\bar{b}$ search at 14 TeV 300$\textrm{fb}^{-1}$}
The most important backgrounds are $Zb\overline{b}$ and $t\overline{t}$. We use {\tt MadGraph5} to generate both signal and background and perform hadronization and the detector simulation with PYTHIA8 and {\tt Delphes3}. The $A\to Zh\to \ell^+ \ell^-  b\bar{b}$ benchmark cross section we used here is 50 fb. 

 \begin{figure}[h]
 \centering
\includegraphics[width=0.45\textwidth]{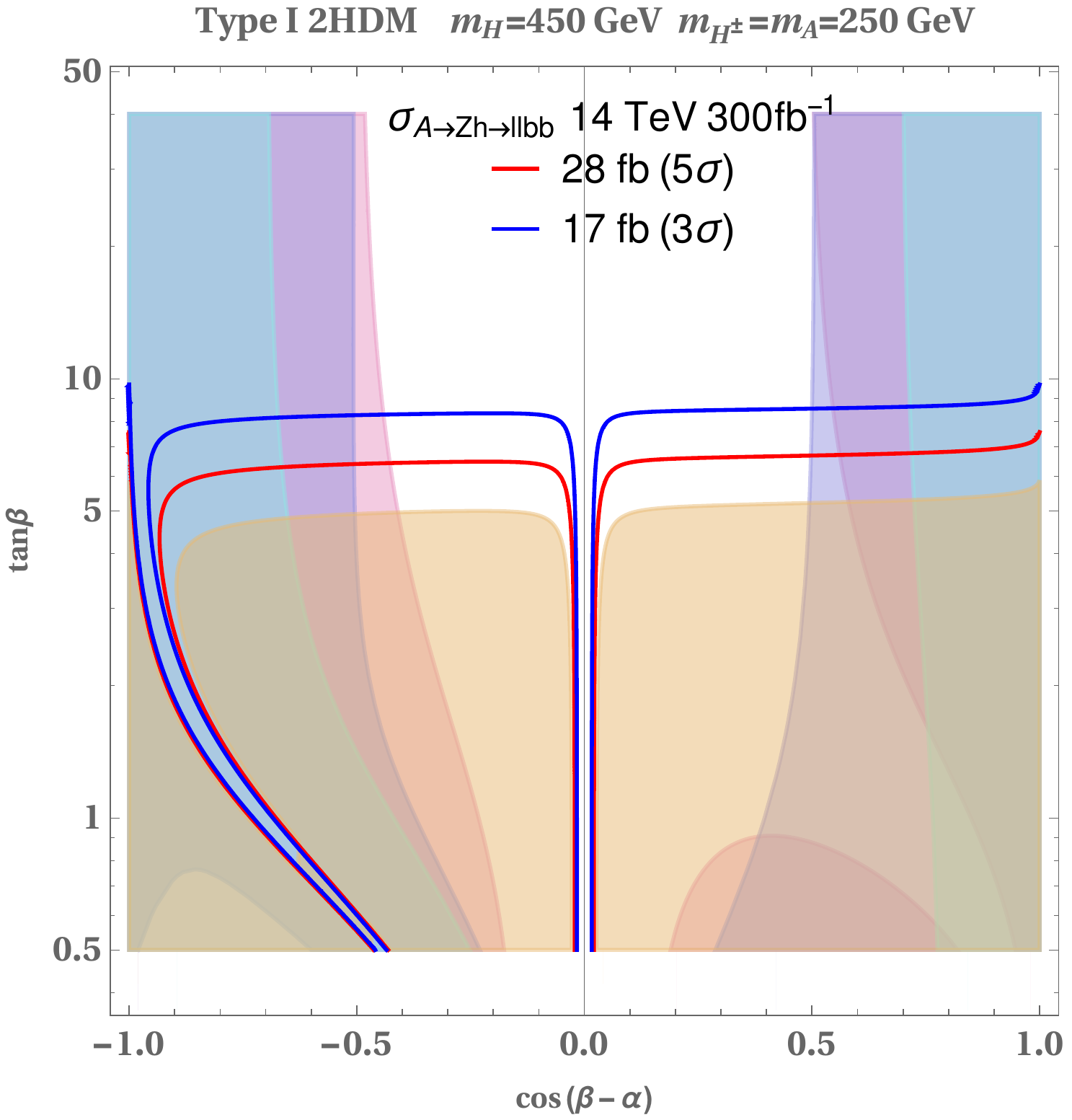}
\caption{$gg\to A\to Zh\to b\overline{b}\ell^+ \ell^- $.}
\label{AZh}
\end{figure}

We are interested in one OSSF lepton pair and two $b$-tagged jets. 
Our requirements are as follows:
\begin{compactitem}
\item Two OSSF leptons with $\pt > 10$ \GeV, $|\eta|<2.5$
\item Lepton isolation: $\Delta R < 0.4$, 
$p_{{\rm T}j}/p_{{\rm T}e} < 0.15$, $p_{{\rm T}j}/p_{{\rm T}\mu} < 0.12$
\item $b$-tagging: fake rate = 0.015, efficiency =  0.75
\end{compactitem}
After preselection, we then optimize the cuts by maximizing $S/\sqrt{B}$. The final selections include
\begin{compactitem}
\item $m_{b\bar{b}} \in [95,135]~\GeV$ 
\item ${\pt}_{b\bar{b}} < 85~\GeV$
\item $m_{\ell^+\ell^-} \in [81, 101]~\GeV$ 
\item ${\pt}_{\ell^+\ell^-} < 85~\GeV$
\item $\met < 95~\GeV$
\item $m_{\ell^+\ell^-b\bar{b}} \in [220, 280]~\GeV$
\end{compactitem}

\begin{table}[h]
\centering
\begin{adjustbox}{max width=\textwidth}
\begin{tabular}{| l | l | l | l |}
    \hline
    event & $\si$(pb) & cut efficiency & yield ($ \mathcal{L}=300\textrm{fb}^{-1}$) \\ 
    \hline
    $Zb\overline{b}$  & 6.7 & $1.0\%$ & $2.0\times 10^4$\\ 
    $t\overline{t}$  &23 & $0.04\%$ & 2500\\
    \hline
    signal type I & 50 & $8.5\%$ & 1300\\
    \hline
\end{tabular}
\end{adjustbox}
\end{table}

We superimpose the cross section contours on top of the experimentally excluded regions from 8 TeV run, 
see Fig.~\ref{AZh}.

\appendix{Appendix C: $H\to ZZ\to \ell^+ \ell^- \ell^+ \ell^- $ search at 14 TeV 300$\textrm{fb}^{-1}$}

The dominant background is $ZZ$. We use {\tt MadGraph5} to generate both signal and background and perform hadronization and the detector simulation with PYTHIA8 and {\tt Delphes3}. The $H\to ZZ\to \ell^+ \ell^- \ell^+ \ell^- $ benchmark cross section is set to be 0.875 fb. 
 \begin{figure}[H]
 \centering
\includegraphics[width=0.45\textwidth]{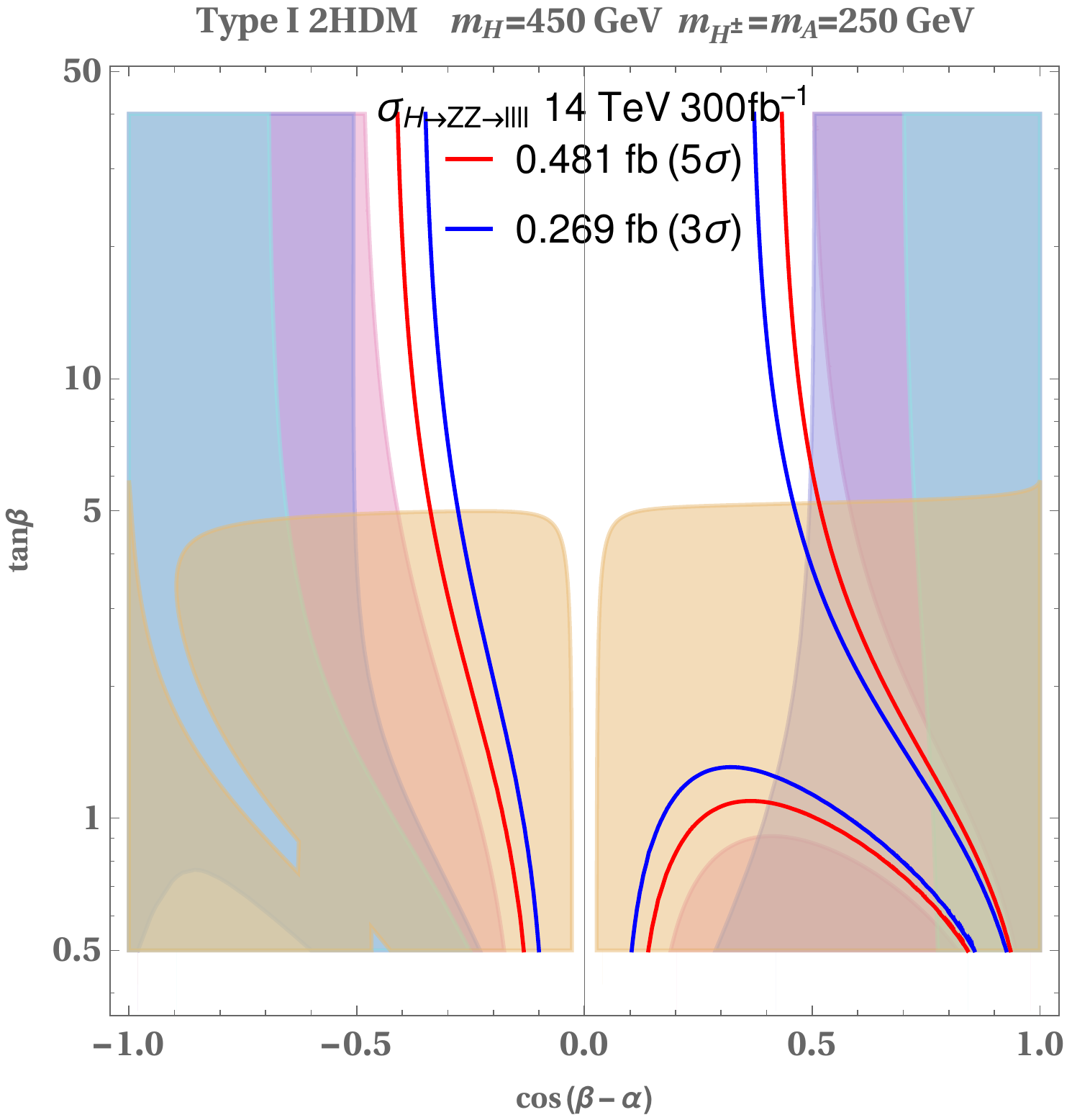}
\caption{$gg\to H\to ZZ\to \ell^+ \ell^- \ell^+ \ell^- $}
\label{HZZ}
\end{figure}
We are interested in 2 pairs of OSSF leptons. We require that
\begin{compactitem}
\item Two OSSF lepton pairs with $\pt > 10$~\GeV, $|\eta|<2.5$
\item The three highest-pT leptons must satisfy $\pt> 20, 15, 10$~\GeV
\item Lepton isolation: $\Delta R < 0.2$, $ {\pt}_j/{\pt}_{\ell} < 0.15$
\end{compactitem}

\noindent
After preselection, we then optimize the cuts by maximizing $S/\sqrt{B}$. The final selections include
\begin{compactitem}
\item Two reconstructed $Z$'s: $m_{\ell^+\ell^-} \in [55, 105]~\GeV$
\item $\Delta\phi$ between leptons in the same OSSF pair $< 2$
\item $m_{\ell^+\ell^-\ell^+\ell^-} \in [365, 475]~\GeV$
\end{compactitem}

\begin{table}[h]
\begin{adjustbox}{max width=\textwidth}
\centering
\begin{tabular}{| l | l | l | l |}
    \hline
    event & $\si$(fb) & cut efficiency & yield ($ \mathcal{L}=300\textrm{fb}^{-1}$) \\ 
    \hline
    $ZZ$  & 33.4 & 0.8\% & 80\\ 
    \hline
    signal type I & 0.875 & 19.8\% & 52 \\
    \hline
\end{tabular}
\end{adjustbox}
\end{table}

\noindent
We superimpose the cross section contours on top of the experimentally excluded regions from 8 TeV run
in Fig.~\ref{HZZ}.


\clearpage

\begin{thebibliography}{90}


\bibitem{Aad:2012tfa}
G.~Aad {\em et~al.}, \emph{Observation of a new particle in the search for the
 Standard Model Higgs boson with the ATLAS detector at the LHC}, {\em Phys.
  Lett.}, vol.~B716, pp.~1--29, 2012.

\bibitem{Chatrchyan:2012xdj}
S.~Chatrchyan {\em et~al.}, \emph{Observation of a new boson at a mass of 125 GeV
  with the CMS experiment at the LHC}, {\em Phys. Lett.}, vol.~B716,
  pp.~30--61, 2012.
  
\bibitem{atlas3}
ATLAS Collaboration,
  \emph{Measurements of the Higgs boson production and decay rates and coupling strengths using pp collision data at $\sqrt{s}=$7 and 8 TeV in the ATLAS experiment}, arXiv:1507.04548 [hep-ex].

\bibitem{cms2}
CMS Collaboration,
  \emph{Precise determination of the mass of the Higgs boson and tests of compatibility of its couplings with the standard model predictions using proton collisions at 7 and 8 TeV}, 	arXiv:1412.8662 [hep-ex].

\bibitem{Alves:2011wf} 
  D.~Alves {\it et al.} [LHC New Physics Working Group Collaboration],
  \emph{Simplified Models for LHC New Physics Searches},
  J.\ Phys.\ G {\bf 39}, 105005 (2012)
  doi:10.1088/0954-3899/39/10/105005,
  arXiv:1105.2838 [hep-ph].



\bibitem{Gunion:1989we} 
  J.~F.~Gunion, H.~E.~Haber, G.~L.~Kane and S.~Dawson,
  \emph{The Higgs Hunter's Guide},
  Front.\ Phys.\  {\bf 80}, 1 (2000).
  
\bibitem{Galloway:2013dma} 
  J.~Galloway, M.~A.~Luty, Y.~Tsai and Y.~Zhao,
  \emph{Induced Electroweak Symmetry Breaking and Supersymmetric Naturalness},
  Phys.\ Rev.\ D {\bf 89}, no. 7, 075003 (2014)
  doi:10.1103/PhysRevD.89.075003,
  arXiv:1306.6354 [hep-ph];
\bibitem{Chang:2014ida} 
  S.~Chang, J.~Galloway, M.~Luty, E.~Salvioni and Y.~Tsai,
  \emph{Phenomenology of Induced Electroweak Symmetry Breaking},
  JHEP {\bf 1503}, 017 (2015)
  doi:10.1007/JHEP03(2015)017,
  arXiv:1411.6023 [hep-ph].

\bibitem{Alves:2012fx} 
  D.~S.~M.~Alves, P.~J.~Fox and N.~Weiner,
  \emph{Supersymmetry with a sister Higgs boson},
  Phys.\ Rev.\ D {\bf 91}, 055003 (2015)
  doi:10.1103/PhysRevD.91.055003,
  arXiv:1207.5522 [hep-ph].

\bibitem{madgraph}
J. Alwall, R. Frederix, S. Frixione, V. Hirschi, F. Maltoni, O. Mattelaer, H.-S. Shao, T. Stelzer, P. Torrielli, M. Zaro,
\emph{The automated computation of tree-level and next-to-leading order differential cross sections, and their matching to parton shower simulations}, 	arXiv:1405.0301 [hep-ph].

\bibitem{pythia}
Torbjörn Sjöstrand, Stephen Mrenna, Peter Skands,
\emph{A Brief Introduction to PYTHIA 8.1},  	arXiv:0710.3820 [hep-ph].

\bibitem{delphes}
J. de Favereau, C. Delaere, P. Demin, A. Giammanco, V. Lemaître, A. Mertens, M. Selvaggi,
\emph{DELPHES 3, A modular framework for fast simulation of a generic collider experiment}, 	arXiv:1307.6346 [hep-ex].

\bibitem{cms3}
CMS Collaboration,
  \emph{Identification of $b$-quark jets with the CMS experiment},   arXiv:1211.4462 [hep-ex].

\bibitem{cms4}
CMS Collaboration,
  \emph{CMS Physics Analysis Summary},  CMS PAS BTV-13-001.

\bibitem{cms5}
CMS Collaboration,
\emph{Electron Reconstruction and Identification at $\sqrt{s} = 8 TeV$},  CMS DP -2010/032.

\bibitem{cms6}
CMS Collaboration,
  \emph{ Electron performance with 19.6 $fb^{-1}$ of data collected at $\sqrt{s} = 8 TeV$ with the CMS detector},  CMS DP -2013/003.

\bibitem{cms7}
CMS Collaboration,
  \emph{Muon Identification performance: hadron mis-Identification measurements and RPC Muon selections},  CMS DP -2014/018.

\bibitem{atlas6}
ATLAS Collaboration,
\emph{Fiducial and differential cross sections of Higgs boson production measured in the four-lepton decay channel in pp collisions at $\sqrt{s}=8$ TeV with the ATLAS detector}, 	arXiv:1408.3226 [hep-ex].

\bibitem{atlas7}
ATLAS Collaboration,
\emph{Search for new light gauge bosons in Higgs boson decays to four-lepton final states in pp collisions at $\sqrt{s}=8$ TeV with the ATLAS detector at the LHC}, 	arXiv:1505.07645 [hep-ex].

\bibitem{cms8}
CMS Collaboration,
\emph{Performance of photon reconstruction and identification with the CMS detector in proton-proton collisions at $\sqrt{s} = 8$ TeV}, 	arXiv:1502.02702 [physics.ins-det].

\bibitem{atlas4}
ATLAS Collaboration,
  \emph{Projections for measurements of Higgs boson cross sections, branching ratios and coupling parameters with the ATLAS detector at a HL-LHC}, ATL-PHYS-PUB-2013-014.

\bibitem{haber10}
Howard E. Haber, Deva O'Neil,
 \emph{Basis-independent methods for the two-Higgs-doublet model III: The CP-conserving limit, custodial symmetry, and the oblique parameters S, T, U}, arXiv:1011.6188 [hep-ph].


 \end{thebibliography}
\end{document}